\documentclass[sigconf, authorversion, nonacm]{acmart}

\acmDOI{10.1145/3658644.3690231}

\usepackage{tikz}
\usepackage{subfigure}
\usepackage{multirow}
\usepackage{amsmath}
\usepackage{lipsum}
\usepackage{threeparttable}
\usepackage{makecell}
\usepackage{titlesec}
\usepackage[ruled,vlined, linesnumbered]{algorithm2e}
\usepackage{algorithmic}
\usepackage{cleveref}
\usepackage{listings}
\usepackage[strict]{changepage}
\usepackage{float}
\usepackage{ulem}
\usepackage{paralist}
\usepackage{threeparttable}
\usepackage{ifthen} 
\usepackage{balance}

\usepackage{url}

\newboolean{showrevisions}
\setboolean{showrevisions}{false} 

\ifbool{showrevisions}
    {\newcommand{\newtext}[1]{\textcolor{blue}{#1}}\newcommand{\removedtext}[1]{\textcolor{red}{\sout{#1}}}}
    {\newcommand{\newtext}[1]{#1}\newcommand{\removedtext}[1]{}}
\begin{document}

\title[ProphetFuzz: Fully Automated Prediction and Fuzzing of High-Risk Option Combinations \\with Only Documentation via Large Language Model]{ProphetFuzz: Fully Automated Prediction and Fuzzing of High-Risk Option Combinations with Only Documentation via Large Language Model}

\author{Dawei Wang}
\orcid{0009-0006-0656-9697}
\affiliation{%
  \institution{Zhongguancun Laboratory}
  \city{Beijing}
  \country{China}
}
\email{wangdw@zgclab.edu.cn}

\author{Geng Zhou}
\orcid{0009-0001-0010-9415}
\affiliation{%
  \institution{Zhongguancun Laboratory}
  \city{Beijing}
  \country{China}
}
\email{zhougeng@zgclab.edu.cn}

\author{Li Chen}
\orcid{0000-0002-4228-7885}
\authornote{Corresponding author.}
\affiliation{%
  \institution{Zhongguancun Laboratory}
  \city{Beijing}
  \country{China}
}
\email{lichen@zgclab.edu.cn}

\author{Dan Li}
\orcid{0000-0002-7581-8865}
\affiliation{%
  \institution{Tsinghua University}
  \city{Beijing}
  \country{China}
}
\email{tolidan@tsinghua.edu.cn}

\author{Yukai Miao}
\orcid{0000-0002-6401-5979}
\affiliation{%
  \institution{Zhongguancun Laboratory}
  \city{Beijing}
  \country{China}
}
\email{miaoyk@zgclab.edu.cn}
\begin{abstract}

Vulnerabilities related to option combinations pose a significant challenge in software security testing due to their vast search space. Previous research primarily addressed this challenge through mutation or filtering techniques, which inefficiently treated all option combinations as having equal potential for vulnerabilities, thus wasting considerable time on non-vulnerable targets and resulting in low testing efficiency. In this paper, we utilize carefully designed prompt engineering to drive the large language model (LLM) to predict high-risk option combinations (i.e., more likely to contain vulnerabilities) and perform fuzz testing automatically without human intervention. We developed a tool called ProphetFuzz and evaluated it on a dataset comprising 52 programs collected from three related studies. The entire experiment consumed \newtext{10.44} CPU years. ProphetFuzz successfully predicted 1748 high-risk option combinations at an average cost of only \$8.69 per program. Results show that after 72 hours of fuzzing, ProphetFuzz discovered 364 unique vulnerabilities associated with 12.30\% of the predicted high-risk option combinations, which was 32.85\% higher than that found by state-of-the-art in the same timeframe. Additionally, using ProphetFuzz, we conducted persistent fuzzing on the latest versions of these programs, uncovering 140 vulnerabilities, with \newtext{93} confirmed by developers and \newtext{21} awarded CVE numbers.

\end{abstract}

\maketitle

\section{Introduction}
\label{sec:introduction}
The growth in software complexity has led to a significant increase in the number of program options and combinations. For example, ImageMagick has 305 options, leading to $2^{305}$ possible combinations, ignoring the different values that each option's parameters can take. This exponential increase in combinations creates many unique execution paths, expanding the search space for vulnerabilities. These paths are only active with specific combinations, challenging traditional mutation-based fuzzing techniques.

In recent years, there has been continuous effort to develop fuzzing methods focused on exploring option combinations. POW\-ER~\cite{lee2022power} and ConfigFuzz~\cite{zhang2023configfuzz} increase path coverage by mutating the options and their values under test. CarpetFuzz~\cite{wang2023carpetfuzz} aims to traverse the complete list of fuzzing combinations by initially filtering out all invalid combinations of options. These efforts have proven to be more effective in testing program paths related to option combinations than traditional fuzzers (e.g., AFL++~\cite{andrea2020afl++}). However, they are still limited in the following aspects: 
\begin{inparaenum}[\bf L1-]
    \item \textit{Lack of Prioritization}: These approaches test each generated combination equally, yet the risk of vulnerabilities varies among different combinations. For example, combinations that involve file parsing options are more likely to have vulnerabilities than combinations that affect only the user interface settings. Excessively testing low-risk combinations (i.e., unlikely to contain vulnerabilities) will reduce the efficiency of fuzz testing.
    \item \textit{Semantic Mismatch}: Every component of an executable command, such as options, values, and files, is interconnected, requiring adherence not just to the correct format and structure (syntactic correctness), but, more importantly, to semantic matching and collaboration to ensure the command executes as intended. Consider the option \textit{``-f avi''}, which specifies that the input must be in AVI format; accordingly, the command with this option functions properly only with inputs in AVI format, and using any other format leads to execution errors. However, in existing proposals, the components of generated commands originate primarily from random mutations or simplistic selections, leading to a lack of coherent and meaningful interaction between components, thereby causing semantic mismatches.
    \item \textit{Reliance on Expertise}: These approaches require extensive manual effort in the setup stages, such as assembling commands, assigning values, and collecting input files, with a strong dependence on the deep expertise of software testers. This not only limits the methods' ability to scale for comprehensive testing, but also raises the barrier of entry.
\end{inparaenum}

If it were possible to incorporate the expertise of security testing experts into an automated tool for predicting high-risk option combinations and configuring fuzz tests, it could address all the limitations mentioned above. This is because reports indicate that experienced security professionals are often able to predict high-risk targets by analyzing documentation and crafting appropriate commands and fuzzing configurations based on the semantic context of these targets~\cite{votipka2018hackers, nosco2020industrial, akgul2023bug}. 

We believe that the recent advancement of large language models (LLMs) has made this idea possible. Trained on extensive text datasets with up to 500 billion tokens~\cite{brown2020language}, these models grasp the complex structures and patterns of language, amassing a wide range of knowledge and the capacity for text comprehension and generation~\cite{achiam2023gpt, touvron2023llama, team2023gemini, anthropic2024claude}. This equips them for a range of sophisticated language understanding and generation tasks. Recent studies show that LLMs can detect hidden patterns in text and even surpass experts in some predictive tasks~\cite{xu2024mental, luo2024large}. Motivated by these findings, we sought to apply LLMs to address the limitations described here. 

However, our initial efforts reveal that direct application of these models did not achieve our expected results (discussed in \newtext{Section~\ref{subsec:evaluation_out-of-the-box}}), which presented the following challenges:

\noindent \textbf{Challenges.} 
\begin{inparaenum}[\bf C1:]
    \item LLMs tend to identify option combinations that violate constraints (i.e., conflicts and dependencies) as high-risk targets, leading to premature program termination and hindering the exploration of program paths. Therefore, informing the LLM of the correct constraints in advance is essential to prevent the generation of invalid combinations. While CarpetFuzz can automatically extract such constraints from documentation using natural language processing (NLP) techniques, its reliance on heuristic rules limits its scalability (discussed in Section~\ref{subsec:evaluation_constraint_extraction}). In contrast, the use of LLMs for constraint extraction offers broader applicability, but also encounters the challenge of LLM ``hallucinations''~\cite{zhang2023siren} (i.e., generating inaccurate or fictitious information). Ensuring the precision of the constraints extracted by LLMs poses a significant challenge.
    \item Although LLMs are adept at understanding and generating natural language, they have not been explicitly trained to recognize the relationships between documentation descriptions and historical high-risk combinations. This gap can lead to a deficiency in specialized knowledge and deep comprehension when handling such tasks. A common strategy to bridge this knowledge gap, known as few-shot learning~\cite{brown2020language}, involves enhancing the model's inference with a few analysis examples. However, the effectiveness of this strategy depends on the availability of such examples, which requires manual analysis by experienced security professionals and is therefore constrained by the availability of expert resources. Consequently, establishing such relationships without direct intervention from experts remains a significant challenge.
    \item To match the semantics of other components within the commands, it may be necessary to produce configuration files and input files in various formats. However, LLMs are primarily skilled in generating text-based inputs. Although the latest multimodal LLMs~\cite{openai2023gpt4v, team2023gemini} can handle inputs such as images, videos, and sounds, they generally do not accommodate the wide range of input formats that different software requires, such as network traffic packets, compressed files, or Executable and Linkable Format (ELF) files. Therefore, effectively enabling models to generate these diverse, non-textual input formats presents a considerable challenge.
\end{inparaenum}

\noindent \textbf{Our Approach.} In this paper, we design ProphetFuzz, an LLM-based, fully automated fuzzing tool for option combination testing. ProphetFuzz can predict and conduct fuzzing on high-risk option combinations~\footnote{\newtext{Due to the nature of fuzzing, this paper focuses on memory corruption vulnerabilities. ``High-risk option combinations'' refer to ``high-risk memory-corruption option combinations'' in the following text.}} with only documentation, and the entire process operates without manual intervention. Given a program's documentation, ProphetFuzz begins by extracting essential details such as the program's name, description, usage synopsis, and option descriptions through keyword matching\newtext{, which is sufficient for this task and cost-effective}. It then leverages the LLM to identify the constraints between options based on these details. To ensure the precision of these identifications, we develop a self-check approach utilizing bidirectional reasoning (C1). \newtext{The intuition behind this approach is that consistent answers from different perspectives suggest higher correctness, similar to methods used in verifying mathematical reasoning~\cite{fu2022complexity,imani2023mathprompter}.} Specifically, this method employs both direct proof and counterproof to assist the LLM in uncovering contradictions in its reasoning, thus allowing for the elimination of inaccuracies in the results identified. Guided by the constraints identified, ProphetFuzz predicts high-risk combinations based on the descriptions of the options. To address the knowledge gap regarding the relationship between documentation and historical high-risk combinations, we design \newtext{a few-shot learning method to bridge these gaps and} an automated few-shot corpus generation method \newtext{to create analysis examples as supplementary knowledge for the LLM} (C2). This method creates eight analysis examples from 29 historical high-risk combinations across eight different programs. Note that the creation of these examples does not require expert participation. For the predicted high-risk combinations, ProphetFuzz initially guides the LLM to comprehend the collective semantics of the option combination and leverage this understanding to allocate option values and files with consistent semantics, thereby generating semantically matched commands. Specifically for configuration and input files, instead of directly generating these files, ProphetFuzz creates Python scripts capable of generating these files and executes these scripts within a sandbox environment to obtain the required files (C3). Finally, ProphetFuzz inputs all generated commands and corresponding files into the fuzzer to carry out fuzz testing on the high-risk option combinations.

We implement a prototype of ProphetFuzz on GPT-4 Turbo, which is currently recognized as the most powerful LLM~\cite{chatbotleaderboard}. We conduct a thorough evaluation of ProphetFuzz across 52 open-source programs collected from datasets used in three previous studies. Within these programs, ProphetFuzz predicted 1748 high-risk option combinations, assembling 7614 commands, and discovered 364 unique vulnerabilities during 72 hours of fuzzing. Of these combinations, 12.30\% were successfully associated with the vulnerabilities. This entire process required an average cost of just \$8.69 per program. Compared to the current state-of-the-art (SOTA) tool, CarpetFuzz, ProphetFuzz utilized fewer commands (0.2$\times$) yet identified more unique vulnerabilities (1.3$\times$), including 224 unique vulnerabilities that CarpetFuzz failed to find. During the process, ProphetFuzz extracted 633 constraints from the documentation with an overall precision of 94.00\% and an average precision of 92.48\%, outperforming CarpetFuzz, which extracted 447 constraints with 76.73\% and 69.72\%. Furthermore, our ablation studies on the option values and files generated by ProphetFuzz indicated that integrating these elements led to the discovery of 34.65\% and 17.24\% more unique vulnerabilities, respectively. We employ ProphetFuzz to perform persistent fuzzing on the latest versions of these programs. To date, ProphetFuzz has uncovered 140 zero-day or half-day~\footnote{Vulnerabilities that have been exposed but have not yet been patched.} vulnerabilities, \newtext{93} of which have been confirmed by the developers, earning \newtext{21} CVE numbers. Investigations into other reported issues are ongoing. Lastly, we analyze how ProphetFuzz predicts high-risk combinations and reveal 15 pieces of knowledge for predicting high-risk combinations from documentation descriptions.

\noindent \textbf{Responsible Disclosure}. To prevent malicious exploitation, all vulnerabilities discovered in this study were promptly disclosed to relevant developers and the CVE organization. So far, \newtext{26} vulnerabilities have been addressed and rectified.

\noindent \textbf{Contributions}. Our contributions are summarized as follows:

$\bullet$\textit{New technique}. We introduce a novel, LLM-based technique for fully automated prediction and fuzz testing of high-risk option combinations. \newtext{We address several critical challenges faced by out-of-the-box LLMs, such as hallucination and knowledge gaps. To tackle these issues, we develop a self-check technique for error correction, a few-shot learning method to bridge knowledge gaps in predicting high-risk combinations, and an automated method to generate few-shot corpora without expert input.} This technique, leveraging only documentation, predicts high-risk combinations without the need for code information and conducts fuzz testing on these combinations entirely automatically \newtext{through our carefully designed prompts}. The entire process, which includes \newtext{document parsing, constraint extraction, }target prediction, command assembly, file generation, and execution of fuzz tests, requires no manual intervention, with an average cost of only \$8.69 per program.

$\bullet$\textit{Implementation and new findings}. We develop the prototype tool ProphetFuzz and thoroughly evaluate its performance across 52 programs gathered from datasets in three prior studies, an effort that consumed a total of \newtext{10.44} CPU years. ProphetFuzz predicts 1748 high-risk option combinations, identifies 364 vulnerabilities, and discovers 140 unique vulnerabilities in the latest versions of these programs, \newtext{93} of which have been confirmed by the developers, earning \newtext{21} CVE numbers. Moreover, our analysis of ProphetFuzz's prediction process highlights 15 key pieces of knowledge for predicting high-risk option combinations, offering insights that are beneficial for further development in this field. We have open-sourced ProphetFuzz \newtext{and the related datasets} to encourage further exploration and research within the community~\footnote{Available in \url{https://github.com/NASP-THU/ProphetFuzz}\label{fn:repo}}.

\section{Background and Related Work}
\label{sec:background}

\subsection{Option-Aware Fuzzing}
\label{subsec:background_option-aware_fuzzing}

The options define the behavior and functionality of the software, allowing users to customize the software to meet their specific needs. This customization maximizes software efficiency across diverse uses and environments. These options encompass a broad range of scenarios, including network settings, resource allocations, and user interface preferences. Although such flexibility markedly improves software usability and the user experience, it also introduces complexity and amplifies the challenges associated with software testing. Each option can influence the behavior of the software, and the combination of multiple options may result in unexpected behavior or security vulnerabilities, particularly without thorough testing. Therefore, understanding and examining these options and their interplay is vital to ensuring software quality and security.

Option-aware fuzzing is tailor-made for this specific requirement. It considers the impact of software options and their combinations, dynamically adjusting and applying the test option combinations on the fly to explore the coverage of different test cases under various option combinations. However, due to the vast search space of option combinations, thoroughly exploring the coverage under all option combinations is unrealistic. To avoid missing vulnerabilities in any potential option combination as much as possible, researchers have developed two categories of option-aware fuzzing techniques: mutation-based~\cite{bohme2017coverage, wang2020tofu, lee2022power, zhang2023configfuzz} and filter-based~\cite{song2020crfuzz, wang2023carpetfuzz}. Mutation-based option-aware fuzzing techniques introduce changes to option combinations via mutations, with the aim of uncovering unexpected vulnerabilities through randomness. AFLargv~\cite{bohme2017coverage} grants fuzzers the capability to mutate test options by treating specific bytes in test cases as options, but introduces a large number of invalid options. Addressing this issue, Tofu~\cite{wang2020tofu} employs structured mutations to alter the presence of each option, while POWER~\cite{lee2022power} builds on this with three mutation operators to further enhance the exploration of unexpected option combinations. Unlike these methods, ConfigFuzz~\cite{zhang2023configfuzz} also considers the impact of the option values and improves the exploration of combination-related paths by mutating both the options and their values. Unlike mutation-based approaches, filter-based option-aware fuzzing techniques prune the search space by filtering out invalid option combinations. CrFuzz~\cite{song2020crfuzz} introduces a clustering-based validity checker, which determines the validity of input options based on the output of the program. CarpetFuzz~\cite{wang2023carpetfuzz}, on the other hand, employs an NLP-based method to extract constraints between options from the program documentation and filters out invalid option combinations that do not satisfy these constraints. However, the methods above do not account for the varying likelihood of vulnerabilities in different combinations of options and treat all combinations equally. This oversight could lead to excessive testing of low-risk option combinations, thereby reducing the opportunities for high-risk combinations to be tested. As a solution, this paper introduces an option-aware fuzzing technique based on the prediction of high-risk option combinations, effectively enhancing fuzzing efficiency.

\subsection{Large Language Model}
\label{subsec:background_llm}

In recent years, large language models (LLMs) such as OpenAI's GPT-4 Turbo have made remarkable advances in NLP. These models benefit from training on massive text datasets, such as 500 billion tokens~\cite{brown2020language}, which allows them to accumulate extensive knowledge and experience. As a result, they can perform a wide range of complex downstream tasks without the need for fine-tuning, excelling in numerous linguistic tasks~\cite{chatbotleaderboard}. Users can guide LLMs to generate outputs aligned with particular intentions by providing prompts in natural language. A prompt is a concise query or task posed to the model designed to elicit goal-oriented responses or content, which is often phrased as a question, statement, or description. When encountering gaps in knowledge, users can supplement the required information by incorporating a few examples into the prompts, a strategy known as ``few-shot learning~\cite{brown2020language}.'' This strategy has been proven to effectively direct LLMs in solving various complex problems, such as olympiad geometry~\cite{trinh2024solving}.

In the field of fuzzing, inspired by the outstanding capabilities of LLMs in text generation and understanding, researchers have started investigating novel methods for using LLMs to improve fuzzing. Using knowledge from vast text datasets, LLMs can efficiently produce code and structured inputs applicable across various domains, significantly improving fuzzing efficiency and coverage. Tools such as Codamosa~\cite{lemieux2023codamosa}, TitanFuzz~\cite{deng2023large}, FuzzGPT~\cite{deng2024large}, CovRL~\cite{eom2024covrl}, Oliinyk et al.~\cite{oliinyk2024fuzzing}, and Fuzz4All~\cite{xia2024fuzz4all} have demonstrated the potential of LLMs in the generation of programming language code. Moreover, ChatFuzz~\cite{hu2023augmenting} and KernelGPT~\cite{yang2023kernelgpt} have delved into LLM applications to generate format-conforming input and syscalls, respectively. PromptFuzz~\cite{lyu2023prompt} has revealed the potential of LLMs to generate fuzz drivers for testing frameworks such as libfuzzer~\cite{libfuzzer}. While these efforts have showcased the potential of LLMs in understanding and generating text inputs, utilizing LLMs to generate non-text inputs, like PDF or ELF files, remains unexplored. The challenge primarily stems from the complexity and structure of non-textual inputs, which exceed LLMs' current generation capabilities. To address this challenge, we propose a method that leverages LLMs to generate code that produces the required inputs. By executing the generated code, we indirectly create the required inputs, effectively circumventing the limitations of LLMs in generating non-text inputs directly.

Beyond generating text input, LLMs are also employed to extract valuable information from external documents, further optimizing the fuzzing process. For instance, ChatAFL~\cite{meng2024large} utilizes LLMs to analyze RFC documents to extract network protocol specifications, enhancing the exploration efficiency of protocol states and code. In this paper, we study an approach to leverage LLMs to extract option descriptions from program documentation to predict high-risk option combinations and integrate synopsis to automatically construct fuzzing configurations for these high-risk combinations. This approach enables a fully automated fuzzing process, from target prediction and command assembly to file generation and fuzz execution, effectively simplifying the complexity of fuzz testing.

\begin{figure}[htbp]
    \centering
    \includegraphics[width=0.9\columnwidth]{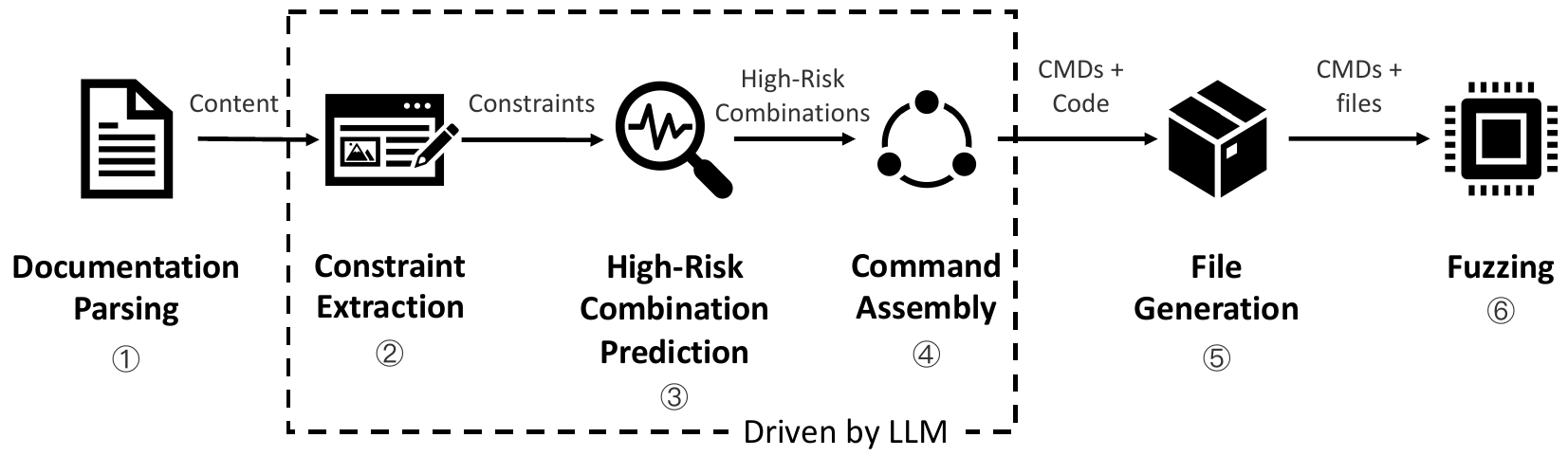}
    \caption{Overview of ProphetFuzz.}
    \label{fig:overview}
\end{figure}
\section{Design}
\label{sec:design}


\subsection{Overview}
\label{subsec:design_overview}

As previously stated, ProphetFuzz relies exclusively on documentation for input. Figure~\ref{fig:overview} illustrates the process in which ProphetFuzz first parses essential documentation content, such as the program description, synopsis, and option details (step 1). \newtext{Considering that document parsing is a simple task, ProphetFuzz employs keyword matching instead of using direct LLM parsing, as keyword matching is both sufficient and cost-effective.} It then identifies the constraints between options and performs a self-check to eliminate erroneous findings (step 2). In the next phase, ProphetFuzz utilizes a few-shot learning approach to predict high-risk option combinations based on the option descriptions while adhering to identified constraints to prevent creating invalid combinations (step 3). Specifically, we manually collect 29 historical high-risk option combinations across eight programs and employ an AutoCoT-based method to generate eight analysis examples as the few-shot (i.e., 8-shot) corpus to assist ProphetFuzz in learning to predict high-risk combinations. Importantly, generating these analysis examples does not require expert participation. ProphetFuzz then constructs the appropriate commands for each identified high-risk combination by selecting suitable option values, guided by the synopsis and option details, and generates Python code to produce the required files (such as configuration and input files) that match these commands semantically (step 4). Finally, ProphetFuzz executes the Python code to produce the files (step 5) and puts them, along with the assembled commands, into the fuzzer for automated fuzzing (step 6).

\begin{figure*}[htbp]
    \centering
    \includegraphics[width=0.9\textwidth]{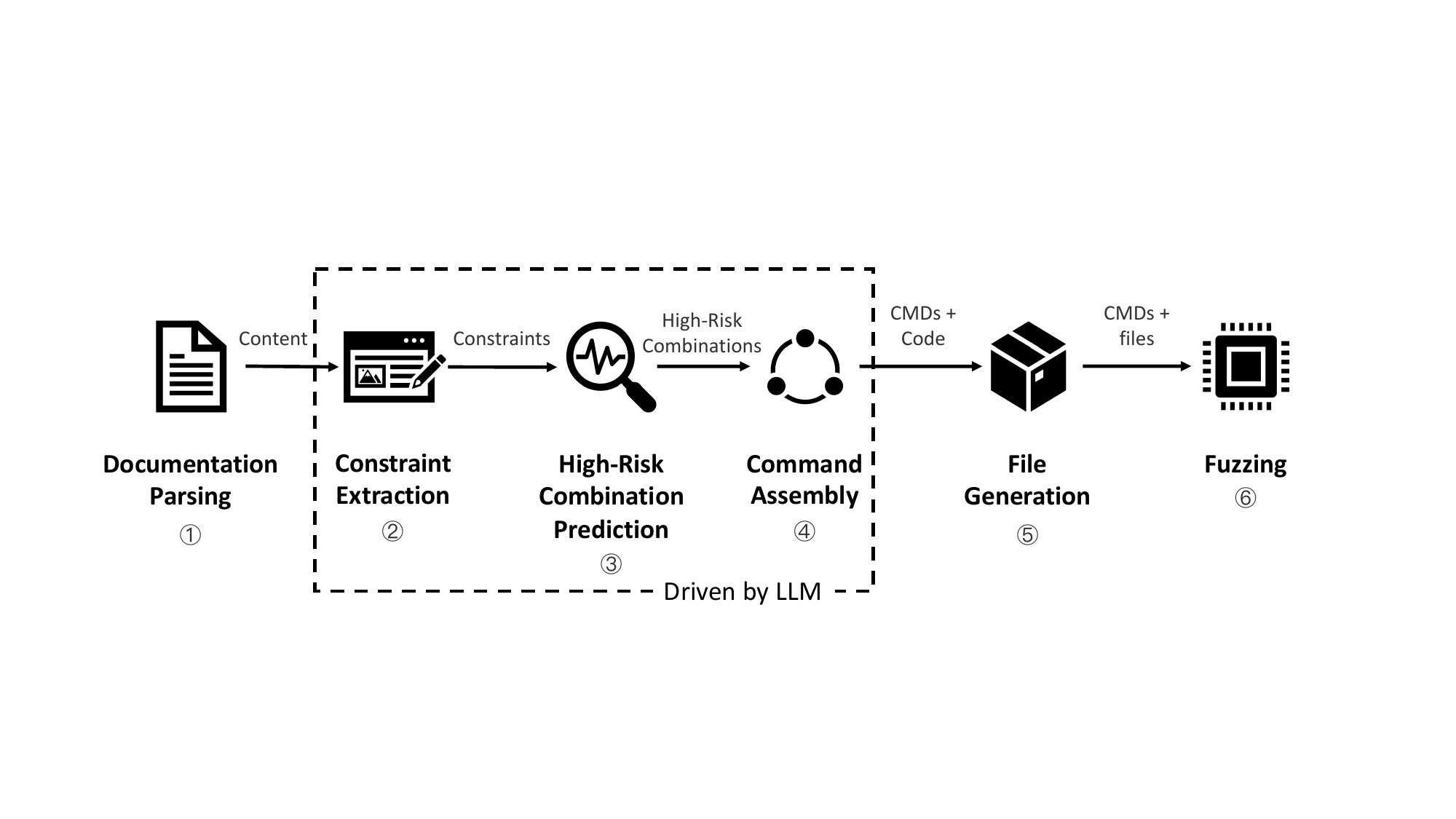}
    \caption{\newtext{Example of ProphetFuzz.}}
    \label{fig:overview_example}
\end{figure*}

\noindent\textbf{\newtext{Example.}} \newtext{Figure~\ref{fig:overview_example} illustrates the process by which ProphetFuzz automatically predicts and executes fuzzing, using the manpage of the well-known video processing software \textit{ffmpeg} as input. Initially, ProphetFuzz extracts the required sections from the \textit{ffmpeg} manpage through keyword matching and stores them as formatted text (JSON) for later use. It then guides the LLM to extract constraints based on option descriptions. Due to the LLM's inherent hallucination issues, some extracted constraints might be incorrect, such as a conflict between \textit{``-start\_at\_zero''} and \textit{``-copyts.''} To address this, ProphetFuzz employs a self-check mechanism to scrutinize all extracted constraints and filter out erroneous results. Based on the document content and constraints, ProphetFuzz predicts high-risk option combinations that do not violate these constraints, such as\textit{``-copyts -start\_at\_zero -y -itsoffset offset -itsscale scale -ss position -sseof position -i url.''} Since no specific values are given for the options, these predicted combinations are not executable. ProphetFuzz subsequently uses the LLM to generate appropriate commands based on the semantic information of the combinations, along with the Python code needed to create the required files. Finally, ProphetFuzz generates the seed corpus by executing the file generation code, modifies the assembled command to be suitable for fuzzing, and executes the fuzzing process.}

\subsection{Constraint Extraction}
\label{subsec:design_extraction}

As noted previously, option combinations that violate constraints between options can cause premature program termination, hindering the fuzzer from exploring paths continuously. To prevent the generation of such combinations, we first utilize LLM to extract constraints between options from the program documentation before predicting high-risk combinations. Specifically, we provide the LLM with the program description and the option descriptions found in the program documentation. We select this information for a couple of reasons. First, to prevent incorrect user configuration of options, authors may have included explicit cues about constraints within this information, such as \textit{``The -alpha flag cannot be used with -mono.''} Secondly, these descriptions also assist LLM in understanding the program's and its options' functionalities, allowing it to infer constraints not explicitly mentioned by the authors based on potential conflicts and dependencies between functionalities. Some option keys, such as \textit{``-r, -R, --recurse-limit, --no-recurse-limit, --recursion-limit, --no-recursion-limit''} may encompass multiple sub-options, which could be different aliases for the same option or entirely distinct options. This ambiguity could affect the LLM's judgment. Consequently, we initially submit keys with more than two sub-options to the LLM for assessment, as this usually indicates that they are distinct options. The prompt is: \textit{``Please separate the options and create individual descriptions for each option based on the original description.''} Following this directive, different sub-options are separated, and their descriptions are adjusted accordingly. These updated data are then provided to the LLM for the subsequent steps.

We then direct the LLM to extract all constraints with the prompt: \textit{``Please find any options that are mutually exclusive or logically conflicting when selected together and find any options that have dependencies on other options.''} We further constrain the LLM's output to a specific format by instructing \textit{``Ensure that the output strictly conforms to JSON format standards, like ...''} to facilitate the automation of subsequent processes. To minimize the risk that the LLM overlooks any constraints, we increase the randomness of the outputs by adjusting the temperature setting of the LLM, which is a sampling parameter ranging from 0 to 1, where higher values indicate a stronger randomness of the output~\cite{openaiapireference}. \newtext{Note that the temperature is an internal hyperparameter of the model, which allows us to make adjustments without modifying the model itself.} Additionally, we instruct the LLM to make multiple inferences for each program and retain the union of the results of each inference to ensure that the extracted constraints are as complete as possible.

\begin{table}[htbp]
    \centering
    \footnotesize
    \caption{Bidirectional questions for conflict and dependency. Here, ``-A'' and ``-B'' represent the subjects of the constraint, indicating either ``-A conflicts with -B'' or ``-A depends on -B.''}
    \begin{tabular}{ccc}
        \hline
        \textbf{Constraint} & \textbf{Type} & \textbf{Question} \\ \hline
        Conflict & Verification & Must -A be used without -B? \\
        Conflict & Counterexample & Can -A be used with -B? \\
        Dependency & Verification & Must -A be used with -B? \\
        Dependency & Counterexample & Can -A be used without -B? \\ \hline
    \end{tabular}
    \label{tab:bidirectional_qustions}
\end{table}

\begin{figure}[htbp]
    \centering
    \includegraphics[width=0.8\columnwidth]{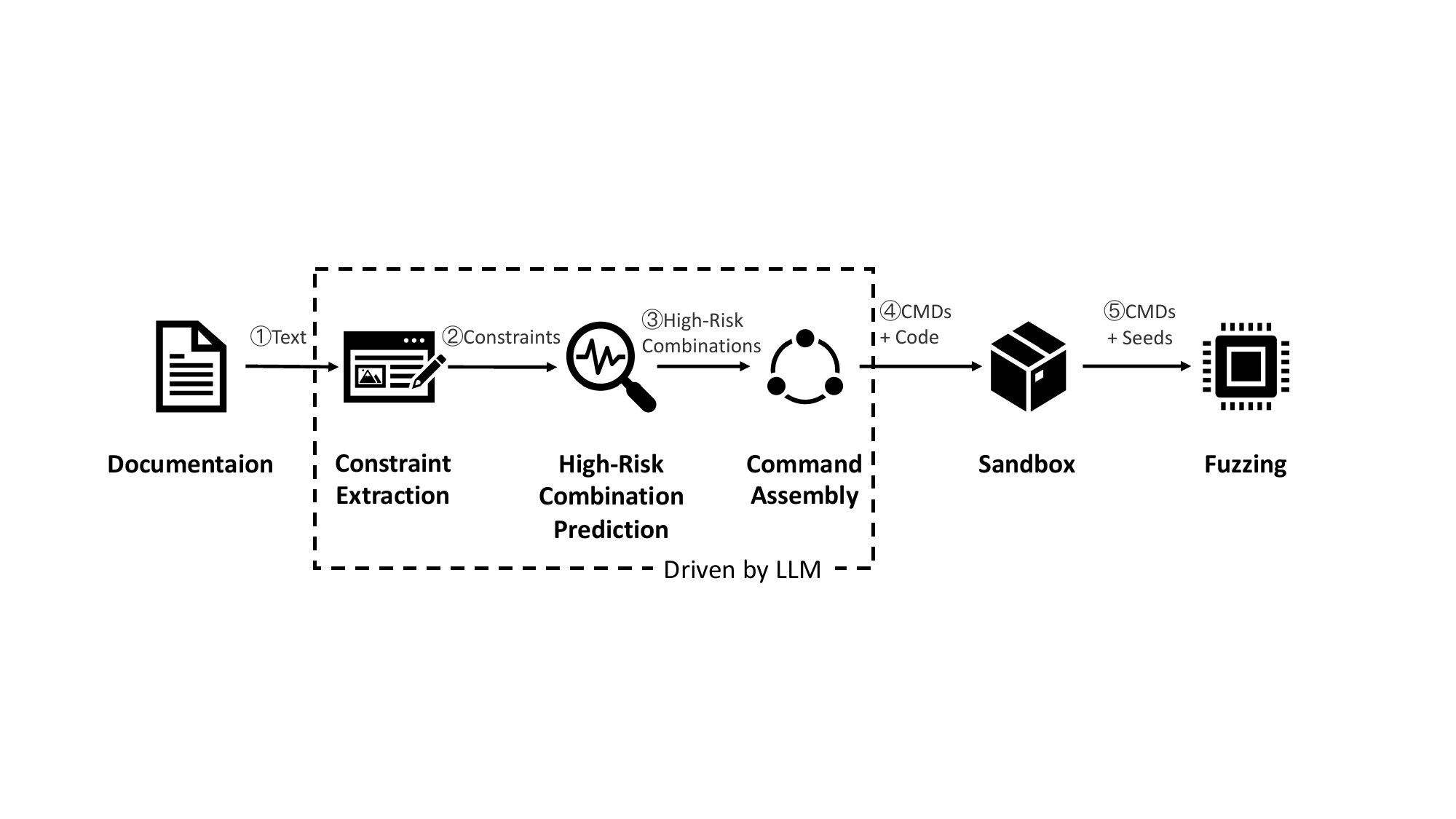}
    \caption{Workflow of the Self-Check approach.}
    \label{fig:self-check}
\end{figure}

However, the inherent hallucinations of LLMs can lead to false positives during constraint extraction, potentially causing high-risk combinations to be mistakenly excluded. Thus, it is crucial to thoroughly validate these constraints, even though the validation process is also susceptible to hallucination problems. In response, we introduce a self-check approach based on bidirectional reasoning. Specifically, for each type of constraint, we formulate a verification question \newtext{based on its specific definition} to confirm its validity and a counterexample question to challenge it, as illustrated in Table~\ref{tab:bidirectional_qustions}. These questions are designed to assess the validity of a constraint from two distinct angles, enabling the LLM to consider the problem comprehensively. The process of this self-check method is shown in Figure~\ref{fig:self-check}. \newtext{After extracting all constraints,} ProphetFuzz automatically generates the corresponding bidirectional questions \newtext{for each constraint based on its type}. Subsequently, these questions, along with descriptions of the options involved, are sent to the LLM for analysis and evaluation, and responses are obtained for each. A constraint is deemed valid if the verification question receives a ``yes'' while the counterexample question receives a ``no.'' To further mitigate the impact of random hallucinations, we adjust the temperature to decrease its randomness and subject each constraint to multiple separate evaluations by the LLM. Each evaluation occurs in a new session to prevent biases from prior communications, enhancing the reliability of the outcomes. A constraint is only considered valid if more than half of the evaluations affirm its correctness. 

The evaluation results in Section~\ref{subsec:evaluation_constraint_extraction} reveal that, with the support of the self-check method, ProphetFuzz extracts 633 constraints from the documentation of 52 programs with an overall precision of \newtext{94.00}\%, and an average precision of \newtext{92.48}\% per program.

\begin{figure*}[htbp]
    \centering
    \includegraphics[width=0.8\textwidth]{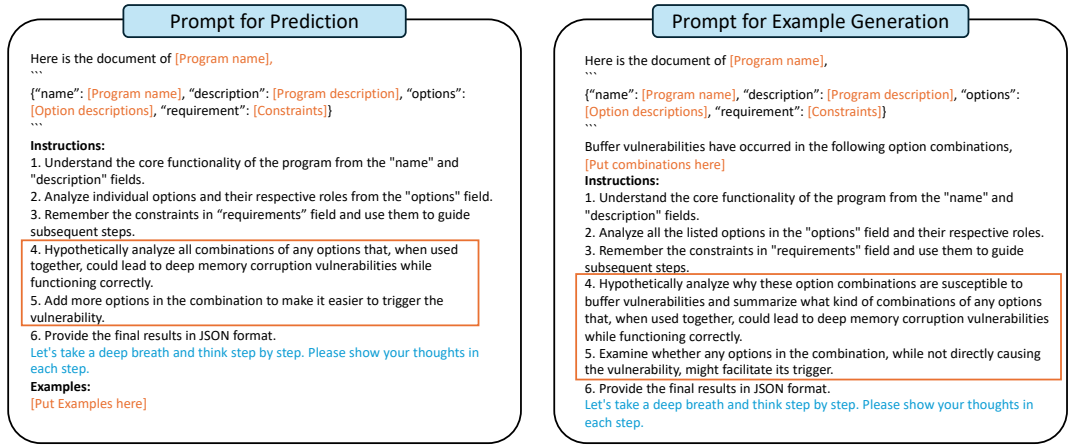}
    \caption{Prompts for prediction and example generation. The boxed area indicates the distinct steps within the two prompts. The content enclosed in brackets denotes the need for specific input.}
    \label{fig:prediction_prompt}
\end{figure*}

\subsection{High-Risk Combination Prediction}
\label{subsec:design_prediction}

Predicting high-risk combinations is a complex task involving multiple steps, including understanding descriptions, traversing combinations, and assessing risk, which poses a challenge to the capabilities of LLMs. A practical solution to this is the use of chain-of-thought (CoT) prompts. A CoT consists of a series of consecutive reasoning steps, which has been proven to significantly improve the ability of LLMs to tackle complex problems~\cite{wei2022chain}. As demonstrated in the prompt for prediction shown in Figure~\ref{fig:prediction_prompt}, we break down the prediction process into six steps using the CoT method to assist the LLM in progressively completing the task. Initially, based on the program name and description, we guide the LLM in understanding the program's core functionality, providing a macroscopic understanding of it (step 1). Next, the LLM analyzes each option and its effects to develop an understanding of the options (step 2). Subsequently, we remind the LLM of the previously extracted constraints to avoid generating invalid combinations that do not comply with these constraints in later predictions (step 3). This step does not involve a specific task for the model; rather, it serves as a reminder to ensure that the model does not forget the constraints identified in earlier steps. Upon completing the preparatory steps above, the LLM examines all combinations and identifies those that may pose vulnerabilities (step 4). During this process, we encourage the LLM to make bold guesses, using the term ``Hypothetically.'' We define high-risk option combinations as \textit{``when used together, they could lead to vulnerabilities in deep memory corruption while functioning correctly.''} This definition emphasizes that the goal of prediction is not to identify errors that cause apparent malfunctions, but to uncover hidden risks that could lead to deep memory corruption. We pay special attention to memory corruption vulnerabilities, as fuzzers primarily target such issues. Experience shows that some options, while not directly causing vulnerabilities, can reduce the difficulty of triggering them. Therefore, once the LLM predicts a combination that includes a memory vulnerability, we further guide the LLM to attempt to add other options that could facilitate triggering the vulnerability (step 5). Finally, we specify that the LLM outputs the results in JSON format, easing the automation of subsequent processes (step 6). Following these six steps, we prompt the LLM with the instruction, \textit{``Let's take a deep breath and think step by step,''} to encourage careful reasoning. This technique has been shown to significantly enhance LLM performance~\cite{yang2023large}. Additionally, we request that the LLM displays its thought process, a practice that ensures that the output of each step provides contextual information for subsequent steps, thereby fostering coherent reasoning by the LLM and increasing the validity of its judgments.

To help the LLM analyze high-risk combinations from the documentation more effectively, we present examples of analysis processes. Ideally, these examples are created manually by human experts (manual-CoT), as expert analyses are most effective in guiding LLMs. However, due to the scarcity of high-level expert resources, we devise an automated method for generating few-shot examples (Auto-CoT) as an alternative. Unlike directly predicting high-risk combinations from documentation, the essence of Auto-CoT lies in directing the LLM to identify potential high-risk factors from the documentation based on historical data. \newtext{To gather this historical data, we review vulnerability reports in GitHub Issues for all popular C/C++ projects with over 100 stars, collecting every related unique combination involving two or more options. From this extensive review, we identify 29 high-risk combinations known to have historical vulnerabilities across eight distinct programs, which represent all the historical high-risk combinations that met our criteria.} Subsequently, we present the LLM with the documentation of these programs and extracted constraints, enabling it to understand program functions and option roles and remember relevant constraints, similar to the initial three steps of the prediction task. To create analysis examples that can serve as few-shot corpus, we modified steps 4-5 in the prediction prompt to ensure the outputs meet our requirements, as illustrated in the boxed area of Figure~\ref{fig:prediction_prompt}. Specifically, starting with historical high-risk combinations, we guide the LLM to hypothetically analyze why these combinations are susceptible to buffer vulnerabilities and summarize which combinations might lead to deep memory corruption vulnerabilities while appearing to function normally. Additionally, we instruct the LLM to explore options that could indirectly facilitate the triggering of vulnerabilities. The outputs from these two steps precisely fulfill the requirements for examples in steps 4-5 of the prediction task. Ultimately, the LLM outputs the final result in JSON format, facilitating our ability to extract and incorporate it into the prediction prompts automatically. Notably, while collecting historical high-risk combinations requires manual effort, it is a one-time job that does not need to be repeated. This approach reduces reliance on human experts and leverages the LLM's robust capabilities to automatically generate helpful analysis examples, enriching the context for predicting high-risk combinations.

Ultimately, ProphetFuzz predicts 1748 high-risk combinations across 52 programs and successfully identifies vulnerabilities in 12.30\% of these combinations, including 364 unique vulnerabilities.

\subsection{Command Assembly}
\label{subsec:design_assembly}

After obtaining the predicted high-risk option combinations, the next step is to assemble these combinations into executable commands for fuzzing. This involves assigning specific values to options and placeholders, setting up necessary configuration files, and specifying input and output files. Typically, this requires manual effort using detailed documentation. To streamline this process, we develop an LLM-based method for automated command assembly. 

Besides providing the program name, description, and option details, we also supply the LLM with the ``synopsis,'' which outlines specific program usage.  Initially, we identify the options to be combined and guide the LLM to generate preliminary executable commands based on this synopsis. We particularly emphasize the need for these commands to adhere to exclusivity, preventing the LLM from concatenating multiple commands with operators like \textit{``\&\&''}, which is unsuitable for fuzzers. The LLM is then tasked with determining which options require values, and cross-verifying this with the synopsis. Next, the LLM needs to understand each command's specific intent and generate valid values for all options that need to be assigned. For other placeholders, we encourage the LLM to creatively assign hypothetical values. To clarify the use of files in commands, we guide the LLM to represent the primary input file as \textit{``file0''} and use sequential placeholders like \textit{``fileN''} for additional necessary files. Since fuzzers typically process only one primary input file at a time, we ensure that ``file0'' is used exclusively in each command, with other required files labeled sequentially from \textit{``file1''} onwards as needed. Finally, we ask the LLM to review each file placeholder to prevent false placeholders from disrupting subsequent processing steps.

After assembling the commands and identifying file placeholders, we further guide the LLM in generating Python code that ``conjures'' files corresponding to each placeholder out of thin air. Initially, we direct the LLM to identify the expected format of each file placeholder, a critical step for selecting the file generation strategy. To avoid incorrect formats, we encourage the LLM to leverage third-party libraries to generate files with complex formats. The LLM then needs to consider the values of other options to determine constraints on the file content, ensuring that the generated files are semantically consistent with the command. Additionally, we specify that the size of \textit{``file0''} should ideally remain under 1KB, while guaranteeing that the command explores as many program behaviors as possible, aligning with fuzzer-recommended practices~\cite{afl_practice}. We have noticed that the LLM occasionally leaves placeholders in the code, expecting users to replace them, such as \textit{``your avi file,''} which can hinder our automated generation process. To address this, we instruct the LLM to replace all placeholders in the code with \newtext{content it deems appropriate}. Finally, we direct the LLM to specify in the generated code that files be saved in the working directory using the given name to facilitate their integration into subsequent automated processes.

In Section~\ref{subsec:evaluation_command_assembly}, we randomly select 25 programs for ablation studies to evaluate the effectiveness of the values and files generated by the command assembly module. The results indicate that with the adoption of these values and files, ProphetFuzz can identify 34.65\% and 17.24\% more vulnerabilities, respectively.

\subsection{File Generation and Fuzzing}
\label{subsec:design_code_execution_and_fuzzing}

Before fuzzing, we execute all generated Python code in a sandbox to create the required files for each command. Next, we compare the generated files with the file placeholders provided by the LLM. If fewer files are generated than placeholders, it suggests that crucial files are missing, making the command unexecutable, and such commands are excluded. However, if the file count matches or exceeds the placeholders but not all placeholders are represented in the generated file list, it indicates that some files have unexpected names. In these cases, we employ a heuristic rule for name correction: if a generated file's name starts with the same prefix as a placeholder, we consider that file to match the placeholder. Commands that cannot be corrected by this rule are also excluded.

For commands that pass this filtering, we further look for placeholders corresponding to input files and replace them with \textit{``@@''} to meet fuzzer specifications. Typically, the placeholder for the input file is \textit{``file0,''} as set in our prompts. For commands deviating from this norm, we search for file placeholders containing key terms like ``input,'' ``test,'' or ``source,'' which are commonly used by LLMs to denote input files. Placeholders with these specific keywords are considered indicative of input files. If a command lacks these keywords but has only one file placeholder, we consider this the input file. Commands failing to meet any of these criteria are excluded from further processing.

After filtering the commands and files, we input them into the fuzzer to initiate fuzz testing. First, we merge all the input files for each program to create an initial corpus. Then, we utilize the corpus minimization tool, a tool designed to refine the initial corpus by eliminating redundant test cases, to produce the minimized corpus tailored to each command and obtain the final minimized corpus for each program by combining them. Finally, we load these corpus and commands into the option-aware fuzzer to begin fuzz testing.
\section{Implementation}
\label{sec:implementation}

We implement a prototype of ProphetFuzz with about 3k lines of Python code, including \newtext{modules for} document parsing, LLM interaction, \newtext{code interpretation,} and fuzzing.

\noindent\textbf{Document Parsing}: We use \textit{Groff} documents, a standard documentation format for command-line programs, as our input. We locate program names using the \textit{``.TH''} control sequence and categorize sections like program descriptions, option descriptions, and synopses with the \textit{``.SH''} control sequence, identifying them by titles such as ``DESCRIPTION'', ``OPTIONS'', and ``SYNOPSIS''. Option descriptions are extracted using sequences like \textit{``.TP''}, \textit{``.PP''}, \textit{``.RS''}, and \textit{``.sp''}. We convert Groff-formatted text into plain, readable natural language using the \textit{``col -b''} command.

\noindent\textbf{LLM Interaction}: We select GPT-4 Turbo (gpt-4-1106-preview) as the LLM backbone for ProphetFuzz, one of the most powerful LLMs currently recognized. Notably, ProphetFuzz's design can adapt to future LLM improvements, enhancing its performance over time. For automation, we interact with the LLM via the OpenAI API~\cite{openaiapireference}. We adjust the LLM's temperature settings to tailor its output~\cite{openaiapireference}: 0.7 for tasks needing diverse outcomes, like constraint extraction and command assembly, and 0.2 for precision tasks like self-checking. For constraint extraction and high-risk option combination prediction, we set the LLM's parameter $n$ to 10, allowing the LLM to make ten inferences each time. When generating few-shot examples, we set $n=1$, as we only need a set of examples. 
During command assembly, to ensure diverse outputs and prevent redundancy, we set $n$ to 3 times the number of options needing assignment ($N$), thus $n=3\times N$.

\noindent\textbf{File Generation and Fuzzing}: Within a Docker container, we set up a Python sandbox using \textit{virtualenv} and pre-install 33 commonly used Python third-party libraries along with 36 command-line tools (for \textit{subprocess} calls) to support the execution requirements of LLM-generated code. These pre-installed libraries and tools have been verified to support the code execution for the 52 programs mentioned in Section~\ref{subsec:evaluation_experiment_setting}. Note that setting up this environment is a one-time job. If there is a need for additional libraries and tools not yet installed, incorporating them is straightforward and easy with the commands \textit{apt} and \textit{pip} to satisfy evolving requirements. For fuzzing, we use \textit{afl-cmin} to minimize the corpus and \textit{CarpetFuzz-fuzzer} as the option-aware fuzzer, which modifies how the target program reads options from command-line to file, facilitating dynamic command adjustments through file changes
\section{Evaluation}
\label{sec:evaluation}

\subsection{Experiment Setting}
\label{subsec:evaluation_experiment_setting}

\noindent\textbf{Platform}. Our experiments are conducted on servers powered by Intel(R) Xeon(R) CPU E5-2630 v3 @ 2.40GHz with 32 cores and 128GB of RAM. No GPU is needed since we do not use local LLM models. These servers run Ubuntu 20.04.

\noindent\textbf{Dataset}. To comprehensively evaluate ProphetFuzz's performance across different types of programs, we integrate datasets from related studies~\cite{lee2022power, zhang2023configfuzz, wang2023carpetfuzz} to form an evaluation dataset, which includes 52 popular open-source programs from 40 packages. The program versions in our dataset align with those specified in the cited studies. The selected programs encompass 26 different input formats, covering text (such as \textit{BNF}, \textit{C code}, \textit{Markdown}, \textit{Mangled Name}, \textit{PostScript}, \textit{JSON}, \textit{ASM}, \textit{PEM}, \textit{SAV}, \textit{TXT}, \textit{XML}, \textit{Rule}), video (\textit{AVI}, \textit{WAV}), audio (\textit{MPG}, \textit{OGG}, \textit{SPX}), images (\textit{JPG}, \textit{EXI}, \textit{GIF}, \textit{PNG}, \textit{TIFF}), and other other binary inputs (\textit{PCAP}, \textit{ELF}, \textit{LRZ}, \textit{PDF}).

\noindent\textbf{Experiment Setup}. We select CarpetFuzz as our comparison benchmark, because it is recognized as the state-of-the-art (SOTA) tool in the field. We initialize its seed corpus using default seed input files sourced from the datasets mentioned above. To ensure fair comparison, both CarpetFuzz and ProphetFuzz are run for equal durations in the same operating environment. Each fuzzing instance is run for 72 hours and repeated five times to mitigate the impact of the inherent randomness associated with fuzzing on the results. Overall, the experiments consume \newtext{10.44} CPU years.

\noindent\textbf{Metrics}. To evaluate the performance of ProphetFuzz, we employ edge coverage and the number of unique vulnerabilities as metrics. Edge coverage is calculated using the \textit{afl-showmap} tool. The identification of unique vulnerabilities is based on the first three entries in call stack reports generated by \textit{ASAN} and \textit{GDB}, as suggested in~\cite{klees2018evaluating}. A unique vulnerability is defined by a distinct combination of the first three entries in its call stack. If two vulnerabilities share the same first three entries, they are considered the same unique vulnerability. Additionally, to highlight the differences in vulnerabilities discovered by ProphetFuzz and CarpetFuzz, we introduce a specific metric for unique vulnerabilities—exclusive vulnerabilities, referring to those found by only one tool within the same timeframe. To represent the upper limit of each tool's capability, we take the union of the results from the five repetitions.

\noindent\textbf{Research Questions}. The following sub-sections explore ProphetFuzz's performance, focusing on these key research questions:

\noindent\textbf{RQ1}: What is ProphetFuzz's end-to-end performance?

\noindent\textbf{RQ2}: How accurate is the constraint extraction module?

\noindent\textbf{\newtext{RQ3}}: \newtext{How do the self-check and few-shot methods contribute to the performance of high-risk option combination predictions?}

\noindent\textbf{\newtext{RQ4}}: \newtext{How does an out-of-the-box LLM perform in predicting high-risk option combinations?}

\noindent\textbf{RQ5}: How does the command assembly module contribute to the overall performance?

\noindent\textbf{RQ6}: What is ProphetFuzz's effectiveness in finding zero-day vulnerabilities?

\noindent\textbf{RQ7}: What knowledge does ProphetFuzz use to predict high-risk option combinations?

\subsection{End-to-end Performance (RQ1)}
\label{subsec:evaluation_end-to-end}

In analyzing the documentation of these 52 programs, ProphetFuzz successfully extracts 643 constraints between options. Based on these constraints, ProphetFuzz further predicts 1748 high-risk option combinations, assembles 7614 unique executable commands, and generates a minimized seed corpus comprising 2656 seed files. We subject these commands and seed files to 72 hours of fuzz testing, with detailed results presented in Table~\ref{tab:end_to_end}.

\begin{table*}[htbp]
    \centering
    \footnotesize
    \caption{Results of ProphetFuzz and CarpetFuzz within 72 hours. The table shows the number of commands (\#Cmds), unique vulnerabilities (\#Uniq Vuls), exclusive vulnerabilities (\#Excl Vuls), the vulnerable combination ratio (Vul Com Ratio), and edge coverage (\#Edge Cov.) for each program. Where $data_P$, $data_C$, and $r$ represent the data of ProphetFuzz, the data of CarpetFuzz, and the ratio between them, respectively.}
    \begin{tabular}{c|cc|cc|cc|cc|ccc}
    \hline
    \multicolumn{1}{c|}{\multirow{2}{*}{\textbf{Program}}} & \multicolumn{2}{c|}{\textbf{\#Cmds}} & \multicolumn{2}{c|}{\textbf{\#Uniq Vuls}} & \multicolumn{2}{c|}{\textbf{\#Excl Vuls}} & \multicolumn{2}{c|}{\textbf{Vul Com Ratio}} & \multicolumn{3}{c}{\textbf{\#Edge Cov.}}\\ \cline{2-12}
        & $data_P$ & $data_C$ & $data_P$ & $data_C$ & $data_P$ & $data_C$ & $data_P$ & $data_C$ & $data_P$ & $data_C$ & $r$ \\ \hline
        avocnv & 497 & 3414 & 4 & \textbf{12} & 2 & \textbf{10} & \textbf{5.45\%} & 2.81\% & \textbf{48951} & 35006 & 139.84\%\\
        bison & 221 & 484 & \textbf{13} & 8 & \textbf{5} & 0 & \textbf{8.82\%} & 0.21\% & 6229 & \textbf{6305} & 98.79\%\\
        c++filt & 25 & 78 & \textbf{96} & 63 & \textbf{60} & 27 & 4.00\% & \textbf{5.13\%} & 3268 & \textbf{3372} & 96.92\%\\
        cflow & 162 & 632 & \textbf{17} & \textbf{17} & \textbf{6} & \textbf{6} & \textbf{14.29\%} & 0.32\% & 1453 & \textbf{1700} & 85.47\%\\
        cjpeg & 53 & 555 & 0 & 0 & 0 & 0 & 0.00\% & 0.00\% & 775 & \textbf{782} & 99.10\%\\
        cmark & 52 & 150 & 0 & 0 & 0 & 0 & 0.00\% & 0.00\% & 7000 & \textbf{7737} & 90.47\%\\
        djpeg & 401 & 1470 & 0 & 0 & 0 & 0 & 0.00\% & 0.00\% & 438 & \textbf{452} & 96.90\%\\
        dwarfdump & 108 & 4704 & 2 & \textbf{4} & 1 & \textbf{3} & \textbf{6.06\%} & 0.23\% & 7767 & \textbf{8042} & 96.58\%\\
        editcap & 276 & 704 & 0 & 0 & 0 & 0 & 0.00\% & 0.00\% & \textbf{4586} & 4408 & 104.04\%\\
        eu-elfclassify & 42 & 1110 & 0 & 0 & 0 & 0 & 0.00\% & 0.00\% & \textbf{213} & 200 & 106.50\%\\
        exiv2 & 224 & 539 & \textbf{2} & \textbf{2} & \textbf{1} & \textbf{1} & \textbf{4.44\%} & 0.56\% & \textbf{8496} & 6657 & 127.63\%\\
        ffmpeg & 336 & 690 & \textbf{4} & 1 & \textbf{3} & 0 & \textbf{7.69\%} & 0.58\% & \textbf{71198} & 62877 & 113.23\%\\
        gif2png & 36 & 263 & \textbf{10} & \textbf{10} & \textbf{1} & \textbf{1} & \textbf{75.00\%} & 11.41\% & \textbf{441} & 434 & 101.61\%\\
        gm & 292 & 1096 & 0 & 0 & 0 & 0 & 0.00\% & 0.00\% & 10172 & \textbf{10588} & 96.07\%\\
        gs & 104 & 78 & 0 & 0 & 0 & 0 & 0.00\% & 0.00\% & \textbf{22747} & 22555 & 100.85\%\\
        img2sixel & 210 & 611 & 5 & \textbf{7} & 3 & \textbf{5} & \textbf{25.00\%} & 1.80\% & 2207 & \textbf{2881} & 76.61\%\\
        jasper & 117 & 82 & 0 & 0 & 0 & 0 & 0.00\% & 0.00\% & \textbf{3416} & 2355 & 145.05\%\\
        jpegoptim & 55 & 988 & 4 & \textbf{6} & 0 & \textbf{2} & \textbf{28.00\%} & 1.82\% & 244 & \textbf{287} & 85.02\%\\
        jpegtran & 143 & 414 & 0 & 0 & 0 & 0 & 0.00\% & 0.00\% & 5773 & \textbf{5831} & 99.01\%\\
        jq & 90 & 558 & 1 & \textbf{3} & 0 & \textbf{2} & \textbf{20.00\%} & 0.36\% & \textbf{2616} & 2257 & 115.91\%\\
        lrzip & 93 & 645 & 0 & 0 & 0 & 0 & 0.00\% & 0.00\% & 4069 & \textbf{4597} & 88.51\%\\
        mpg123 & 78 & 763 & 0 & \textbf{4} & 0 & \textbf{4} & 0.00\% & \textbf{3.54\%} & 3043 & \textbf{3157} & 96.39\%\\
        mutool & 211 & 769 & \textbf{1} & \textbf{1} & \textbf{1} & \textbf{1} & \textbf{2.44\%} & 0.13\% & 13118 & \textbf{16238} & 80.79\%\\
        nasm & 238 & 537 & \textbf{17} & 11 & \textbf{10} & 4 & \textbf{21.95\%} & 3.35\% & 7399 & \textbf{7623} & 97.06\%\\
        nm & 63 & 555 & \textbf{13} & 5 & \textbf{10} & 2 & \textbf{3.13\%} & 0.36\% & \textbf{5860} & 4986 & 117.53\%\\
        objdump & 209 & 1742 & \textbf{9} & 3 & \textbf{9} & 3 & \textbf{11.32\%} & 0.11\% & \textbf{11392} & 10793 & 105.55\%\\
        ogg123 & 165 & 257 & 0 & \textbf{1} & 0 & \textbf{1} & 0.00\% & \textbf{0.39\%} & \textbf{385} & 377 & 102.12\%\\
        openssl-asn1parse & 142 & 249 & 0 & \textbf{1} & 0 & \textbf{1} & 0.00\% & \textbf{1.20\%} & \textbf{4415} & 2610 & 169.16\%\\
        openssl-ec & 363 & 386 & 0 & 0 & 0 & 0 & 0.00\% & 0.00\% & \textbf{9474} & 7979 & 118.74\%\\
        openssl-rsa & 194 & 2043 & 0 & 0 & 0 & 0 & 0.00\% & 0.00\% & \textbf{8034} & 7105 & 113.08\%\\
        pdftohtml & 73 & 456 & \textbf{3} & \textbf{3} & 0 & 0 & \textbf{33.33\%} & 0.22\% & 5635 & \textbf{7764} & 72.58\%\\
        pdftopng & 181 & 326 & \textbf{12} & 3 & \textbf{10} & 1 & \textbf{60.71\%} & 1.84\% & 4531 & \textbf{5043} & 89.85\%\\
        pdftops & 122 & 1495 & \textbf{11} & 6 & \textbf{7} & 2 & \textbf{77.14\%} & 0.94\% & \textbf{3978} & 1784 & 222.98\%\\
        pdftotext & 172 & 787 & \textbf{10} & 5 & \textbf{5} & 0 & \textbf{70.59\%} & 1.91\% & \textbf{4485} & 1500 & 299.00\%\\
        pngfix & 63 & 116 & 0 & 0 & 0 & 0 & 0.00\% & 0.00\% & \textbf{1058} & 1015 & 104.24\%\\
        podofoencrypt & 33 & 203 & \textbf{1} & 0 & \textbf{1} & 0 & \textbf{72.73\%} & 0.00\% & \textbf{2796} & 251 & 1113.94\%\\
        pspp & 166 & 258 & \textbf{46} & 19 & \textbf{29} & 2 & \textbf{90.91\%} & 8.14\% & \textbf{15958} & 11005 & 145.01\%\\
        readelf & 106 & 1130 & \textbf{10} & 3 & \textbf{10} & 3 & \textbf{13.33\%} & 6.55\% & \textbf{12305} & 11741 & 104.80\%\\
        size & 94 & 146 & 0 & 0 & 0 & 0 & 0.00\% & 0.00\% & \textbf{3708} & 2894 & 128.13\%\\
        speexdec & 76 & 271 & 0 & 0 & 0 & 0 & 0.00\% & 0.00\% & \textbf{784} & 596 & 131.54\%\\
        tcpprep & 110 & 710 & 2 & \textbf{4} & 0 & \textbf{2} & \textbf{16.67\%} & 4.23\% & 707 & \textbf{902} & 78.38\%\\
        tcpreplay & 65 & 1328 & 1 & \textbf{2} & 0 & \textbf{1} & \textbf{31.82\%} & 2.64\% & 733 & \textbf{765} & 95.82\%\\
        tiff2pdf & 120 & 584 & 0 & 0 & 0 & 0 & 0.00\% & 0.00\% & 6250 & \textbf{6268} & 99.71\%\\
        tiff2ps & 78 & 832 & 0 & 0 & 0 & 0 & 0.00\% & 0.00\% & 3932 & \textbf{4073} & 96.54\%\\
        tiffcp & 45 & 385 & \textbf{3} & \textbf{3} & \textbf{2} & \textbf{2} & \textbf{14.71\%} & 0.52\% & 5710 & \textbf{6474} & 88.20\%\\
        tiffcrop & 334 & 680 & \textbf{65} & 45 & \textbf{47} & 27 & \textbf{26.67\%} & 7.65\% & 6283 & \textbf{6810} & 92.26\%\\
        tiffinfo & 17 & 116 & 0 & 0 & 0 & 0 & 0.00\% & 0.00\% & 3789 & \textbf{3822} & 99.14\%\\
        vim & 243 & 1031 & 0 & \textbf{2} & 0 & \textbf{2} & 0.00\% & \textbf{0.19\%} & 10630 & \textbf{49166} & 21.62\%\\
        xmlcatalog & 66 & 116 & 0 & 0 & 0 & 0 & 0.00\% & 0.00\% & \textbf{6129} & 5864 & 104.52\%\\
        xmllint & 109 & 664 & 2 & \textbf{17} & 1 & \textbf{16} & 2.22\% & \textbf{14.31\%} & \textbf{12609} & 10946 & 115.19\%\\
        xmlwf & 67 & 329 & 0 & 0 & 0 & 0 & 0.00\% & 0.00\% & \textbf{3473} & 3104 & 111.89\%\\
        yara & 74 & 455 & 0 & \textbf{3} & 0 & \textbf{3} & 0.00\% & \textbf{0.22\%} & 2588 & \textbf{3021} & 85.67\%\\ \hline
        \textbf{Count} & 7614 & 38984 & \textbf{364} & 274 & \textbf{224} & 134 & \textbf{12.30\%} & 1.50\% & 393250 & \textbf{394999} & 99.56\%\\ \hline
    \end{tabular}
    \label{tab:end_to_end}
\end{table*}

Within these 52 programs, ProphetFuzz identifies 364 unique vulnerabilities, which is 1.33 times more than that of CarpetFuzz (274 vulnerabilities discovered) under the same conditions.
\newtext{Benefiting from ProphetFuzz's precise capabilities in predicting high-risk option combinations and assembling commands (see Sections~\ref{subsec:evaluation_self-check_few-shot} and~\ref{subsec:evaluation_command_assembly}),} it identifies 224 vulnerabilities that CarpetFuzz fails to detect, representing 61.54\% of all vulnerabilities found, which demonstrates ProphetFuzz's efficiency in vulnerability detection. Note that CarpetFuzz also identifies 134 vulnerabilities that ProphetFuzz misses\newtext{, especially in \textit{avconv}, \textit{c++filt}, \textit{tiffcrop}, and \textit{xmllint}, which accounts for 60\%. We analyze cases from these four programs and identify the following reasons for ProphetFuzz's omissions.}

\newtext{Thirty-seven missed vulnerabilities (7 in \textit{avconv}, 2 in \textit{c++filt}, 27 in \textit{tiffcrop}, and 1 in \textit{xmllint}) arise because ProphetFuzz focuses on predicting high-risk combinations rather than exhaustively exploring all possibilities like CarpetFuzz. When queried, the backend LLM identifies 92.86\% of these combinations as high-risk, indicating ProphetFuzz's potential to predict most of these overlooked combinations. Due to cost and performance considerations, ProphetFuzz consolidates predictions from only ten inferences, which may limit the coverage of potential combinations. Expanding the number of inferences could help reduce these misses.
Thirty-eight vulnerabilities occur because ProphetFuzz, while successfully predicting high-risk combinations, fails to specify the necessary values to trigger these vulnerabilities. Specifically, 15 vulnerabilities in \textit{xmllint} are only triggered when \textit{``–maxmem''} falls within a certain range (see Section~\ref{subsec:evaluation_command_assembly}), and 23 in \textit{c++filt} only occur when \textit{``-s''} specifies a DLang compiler instead of a C++ compiler. While CarpetFuzz may inadvertently meet these conditions through random selection, ProphetFuzz, aiming for validity, typically opts for more commonly used values, such as selecting a C++ compiler for \textit{c++filt}, thus missing these vulnerabilities. Incorporating less common values could help address this issue in the future.
Lastly, four vulnerabilities are missed due to randomness (2 in \textit{avconv}, 2 in \textit{c++filt}). ProphetFuzz successfully predicts and assembles certain commands, yet temporarily fails to detect them.}

Additionally, we analyze the option combinations related to these vulnerabilities. Of the 1748 high-risk option combinations predicted by ProphetFuzz, 12.30\% trigger vulnerabilities within 72 hours, surpassing CarpetFuzz's 1.50\% (8.2$\times$). Specifically, in six programs (\textit{gif2png}, \textit{pdftopng}, \textit{pdftops}, \textit{pdftotext}, \textit{podofoencrypt}, and \textit{pspp}), more than half of the predicted high-risk combinations successfully trigger vulnerabilities, with the ratio reaching as high as 90.91\% in \textit{pspp}. This demonstrates ProphetFuzz's efficiency in testing option combination-related vulnerabilities and highlights the effectiveness and necessity of predicting high-risk combinations.

To further compare the performance of ProphetFuzz with CarpetFuzz, we also measure the coverage achieved by both tools across 52 programs. Despite not primarily aiming to maximize path coverage, ProphetFuzz's coverage is remarkably close to CarpetFuzz's, with a marginal difference of less than 0.5\%. Although ProphetFuzz uses only 0.2 times as many commands as CarpetFuzz, the coverage difference is within 30\% for most programs (43 out of 52). For these programs, 24 exhibit a coverage discrepancy of less than 10\%. Notably, on \textit{vim}, CarpetFuzz discovers 4.62 times more paths than ProphetFuzz. Analysis reveals that in the POWER dataset, the initial command for vim is \textit{``-u NONE -X -Z -e -s -S @@ -C `:qa!'''}, indicating that the fuzzer is mutating the value of the \textit{-S} option (a vim script) rather than the input file. CarpetFuzz includes these options in every command generated. In contrast, the LLM behind ProphetFuzz does not know this. It continues to treat the input file as the mutation target. In the \textit{vim} commands, the majority involve only reading operations on the input file. Regardless of how the input file is mutated, the range of executable operations remains limited, thus covering fewer paths compared to mutations applied to the \textit{vim} script (\textit{-S}). In future research, to address this issue, we plan to allow the LLM to autonomously choose the most suitable file for mutation rather than strictly mutating the input file.

Furthermore, ProphetFuzz outperforms CarpetFuzz (>2$\times$) in coverage on three programs (\textit{pdftotext}, \textit{pdftops}, and \textit{podofoencrypt}), with the coverage on \textit{podofoencrypt} being ten times that of CarpetFuzz. We attribute this to the fact that these programs perform strict checks on input formats, while CarpetFuzz uses inherently corrupted default PDF seeds in the dataset, which are from the AFL project~\cite{pdf_seed} and are widely used in many studies~\cite{wang2023carpetfuzz, pang2023ocfi, zhang2023profile}. These corrupted seeds cause programs to exit early, leading to low coverage. PDF is a highly structured format, and simply mutating it is often insufficient to repair format damage. In contrast, ProphetFuzz instructs the LLM to generate seeds that meet complex format requirements by calling third-party libraries (e.g. \textit{pyFPDF}~\cite{pyfpdf}), achieving higher coverage on these programs. While manually collecting more suitable seeds could mitigate this issue, the diversity of input formats would necessitate considerable manual effort. Additionally, even within the same input format, there can be varying detailed requirements (e.g., styles in PDF), adding complexity to the manual collection of seeds. Instead, ProphetFuzz enhances its ability to meet large-scale testing needs by guiding LLMs to automate the generation of high-quality, specification-compliant seeds. \newtext{To ensure adherence to specific requirements, ProphetFuzz directs LLMs to utilize Python libraries and external tools for file creation. For instance, LLM uses the function \textit{``pdf.set\_font(`Arial',size=12)''} to meet font specifications, thus enabling precise and accurate file generation while eliminating the need for manual script tuning.}

\subsection{Precision of Constraint Extraction (RQ2)}
\label{subsec:evaluation_constraint_extraction}

\begin{figure}[htbp]
    \centering
    \includegraphics[width=0.6\columnwidth]{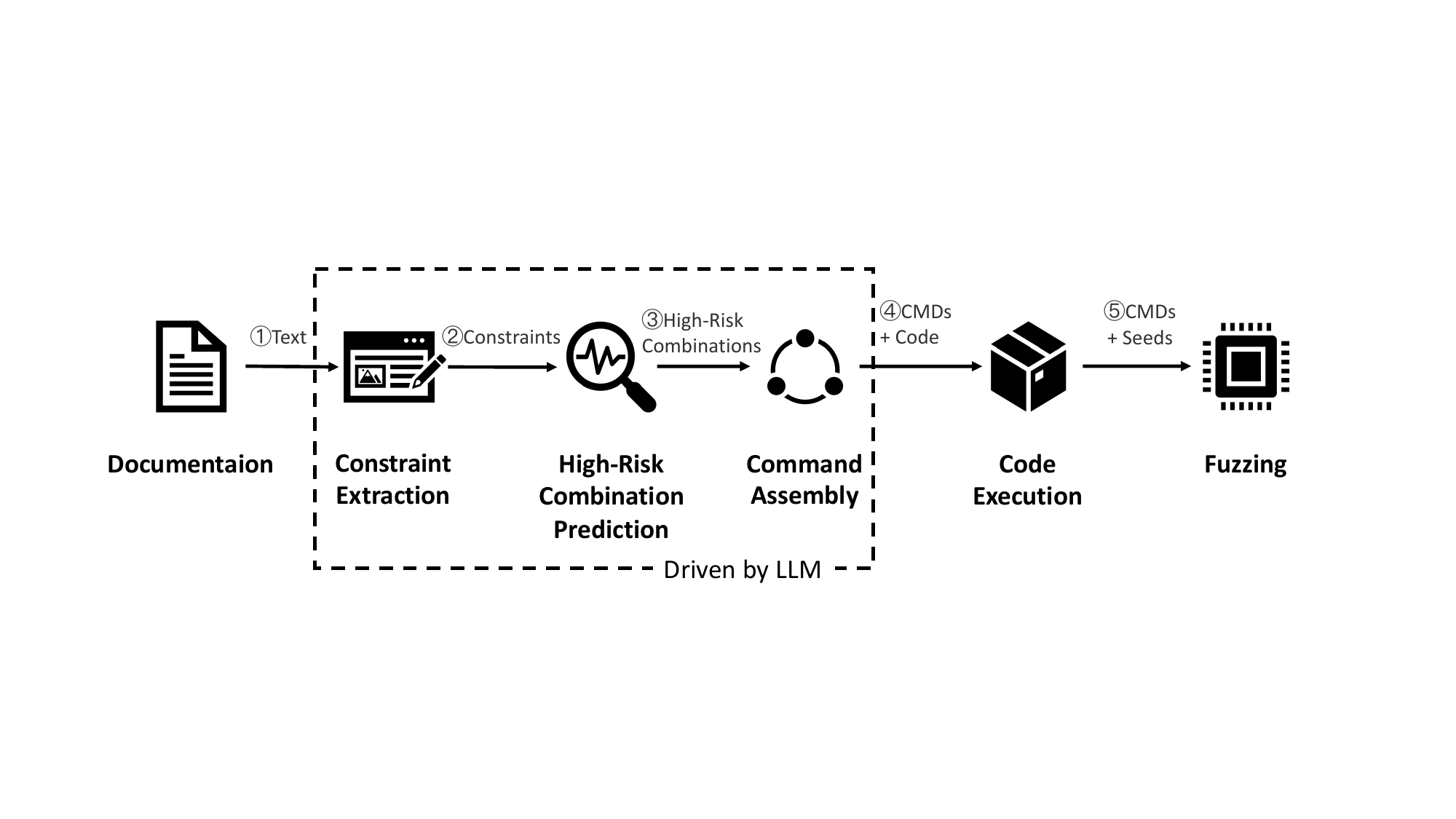}
    \caption{Boxplot comparing the precision of constraint extraction across 52 programs by ProphetFuzz, CarpetFuzz, and ProphetFuzz$^{NSC}$ (without self-check).}
    \label{fig:constraint_precision}
\end{figure}

\newtext{To compare the performance of ProphetFuzz and CarpetFuzz in constraint extraction, we manually annotate all extracted constraints.} Within the documentation of these 52 programs, ProphetFuzz successfully extracts 633 constraints with an overall precision of 94.00\% and an average precision of 92.48\% per program. In contrast, CarpetFuzz extracts only 447 constraints, achieving an overall precision of \newtext{76.06\%} and an average precision of \newtext{68.32\%}, all lower than those of ProphetFuzz. To compare the precision of constraint extraction between these two tools across different programs, we plot their boxplots as shown in Figure~\ref{fig:constraint_precision}. The data range for ProphetFuzz's boxplot is concentrated between 77.78\% and 100\%, with a median of 100\%, indicating that ProphetFuzz maintains high precision in extracting constraints across most programs. On the other hand, the boxplot for CarpetFuzz displays a data spread from 0\% to 100\%, with a median of 75\%, reflecting high variability in performance and less consistency. This difference is due to CarpetFuzz relying on heuristic rules for extracting constraints, whereas ProphetFuzz utilizes carefully designed prompts to guide the LLM in extracting constraints, resulting in better precision and applicability. Note that, the boxplot for ProphetFuzz shows three outliers, which correspond to precision on six programs: \textit{dwarfdump} and \textit{tiffcp} at 50\%, \textit{nasm} at 63.64\%, and \textit{editcap}, \textit{ogg123}, and \textit{xmlcatalog} at 66.67\%. These outliers indicate unusually low constraint extraction precisions for ProphetFuzz on these six programs. Upon analysis, we attribute the issue to misunderstandings by the LLM. Despite our development of a set of self-check questions based on bidirectional reasoning to aid the LLM's comprehension of our constraint definitions, the LLM's inherent randomness lead to incomplete understanding in some instances, resulting in false positives. In five programs except \textit{nasm}, the number of constraints extracted is limited (six or fewer). Hence, each false positive significantly impacts precision. Despite extracting a relatively higher number of constraints (11 in total) from \textit{nasm}, the LLM's misunderstanding of a complex, multi-option constraint leads to a disproportionate increase in false positives when these constraints are split into pairs for individual assessment, subsequently resulting in reduced precision.

As mentioned above, ProphetFuzz employs a bidirectional reason\-ing-based self-check method to inspect and filter out incorrect constraint extraction results. To assess the effectiveness of this method, we analyze the constraint extraction outcomes before applying the self-check (ProphetFuzz$^{NSC}$). Notably, ProphetFuzz$^{NSC}$ extracts 6682 constraints, which is 10.56 times the total number of constraints after implementing self-check. However, these constraints have an overall precision of only 23.41\% and an average precision of 33.37\%, indicating that despite the LLM's strong text comprehension and reasoning, it cannot ensure correct constraint extraction results without our self-check method. We also plot a boxplot for the precision of ProphetFuzz$^{NSC}$ across different programs, as shown in Figure~\ref{fig:constraint_precision}. The data range for ProphetFuzz$^{NSC}$'s boxplot is concentrated between 2.17\% and 80\%, with a median of 23.90\%, indicating generally low and unstable precision across most programs. Additionally, three outliers in the boxplot correspond to four programs with unusually high precisions: \textit{eu-elfclassify} at 88.24\%, \textit{size} at 90.91\%, \textit{c++filt}, and \textit{tcpprep} at 100\%, which suggest that ProphetFuzz$^{NSC}$ performs extraordinarily well on these specific programs. Further analysis reveals that the high precision in these programs is largely due to the precise documentation of constraints. For instance, in \textit{eu-elfclassify}'s document, options related to constraints are described using a consistent sentence structure. Documentation for \textit{c++filt} and \textit{size} groups options with constraints together for clarity, while \textit{tcpprep} explicitly specifies the corresponding constraints within the description of each option. Such structured documentation significantly facilitates the LLM's high precision in extracting constraints, thus enabling these programs to demonstrate higher precision even without the self-check process.

We observe a reduction of 969 true positive (TP) constraints following self-check. Most of these constraints are associated with help and version options such as \textit{``--help''} and \textit{``--version''}. Before the self-check, we increase the temperature setting to encourage the LLM to make bolder guesses. This leads to identifying combinations of these options with others as potential conflicts. Since these options override the functionality of others, we mark them as TPs during manual annotation. However, during the self-check, the LLM's settings are adjusted to favor more cautious outcomes. As a result, it thinks that while these options might override other options' functionalities, this does not necessarily constitute an actual conflict. For example, during the self-check of the conflict constraint between \textit{--debug-level} and \textit{--help} in \textit{jasper}, the LLM states: \textit{``--debug-level can be used with --help, but standard behavior would suggest that the help message will be displayed, and the program will exit without considering the --debug-level option.''} Note that these disregarded constraints do not affect the subsequent prediction of high-risk option combinations, as the LLM does not consider combinations involving such options to present vulnerabilities by default.

\newtext{To further assess the robustness of our constraint extraction module, we randomly select 20 programs and manually annotate their documents, totaling about 22,730 words, to calculate recall (See details in footnote~\ref{fn:repo}). We exclude help and version constraints to ensure reliability and find 180 constraints as ground truth. ProphetFuzz extracts 148 constraints, with a precision of 97.97\% and a recall of 80.56\%. CarpetFuzz extracts 108 constraints with lower precision (77.78\%) and recall (46.67\%), highlighting ProphetFuzz's superior performance. ProphetFuzz$^{NSC}$, evaluated without help and version constraints, extracts 1880 constraints but with only 8.56\% precision; however, its recall is the highest at 89.44\%, reflecting the backend LLM's extensive initial extraction of constraints before self-check, as expected. Currently, we use a self-check voting threshold of 5, where higher thresholds improve precision but reduce recall, and lower thresholds do the opposite. Future research could aim to identify the optimal threshold for peak performance.}

\noindent\textbf{Interesting finding.} Previous work~\cite{wang2023carpetfuzz} mentions four hidden conflicting constraints in \textit{img2sixel}: \textit{-P} vs \textit{-8}, \textit{-p} vs \textit{-e}, \textit{-p} vs \textit{-I}, and \textit{-p} vs \textit{-b}. These conflicts are not documented online and typically only surface through error messages when violated. Despite insufficient indications, ProphetFuzz successfully identifies two constraints (\textit{-p} versus \textit{-e} and \textit{-p} versus \textit{-I}). Our analysis of the LLM's extraction process reveals that ProphetFuzz can identify these hidden constraints based on a nuanced understanding of the options' functionalities. For instance, the LLM infers that \textit{``-p has a default value of 256, which implies it defines the color depth. This might conflict with -e as monochrome means 2 colors.''} This analysis relies on sophisticated reasoning capabilities, which traditional heuristic-based constraint extraction tools generally fail to handle.

\subsection{\newtext{Contribution of Self-Check and Few-Shot Methods (RQ3)}}
\label{subsec:evaluation_self-check_few-shot}

\newtext{This paper introduces a self-check mechanism and a few-shot method to boost ProphetFuzz's predictions. To assess the impact of these methods on prediction performance, we implement two variants of ProphetFuzz: ProphetFuzz$^{NSC}$ (without the self-check mechanism in constraint extraction), and ProphetFuzz$^{ZS}$ (zero-shot, without the few-shot method). We conduct 72-hour fuzz tests on 25 randomly selected programs using both variants and the original ProphetFuzz under the same conditions~\footnote{\newtext{The experiments in Sections~\ref{subsec:evaluation_self-check_few-shot} and~\ref{subsec:evaluation_out-of-the-box} were conducted in a different period than those in Sections~\ref{subsec:evaluation_end-to-end} and~\ref{subsec:evaluation_command_assembly}. To ensure uniform conditions, we reran ProphetFuzz.}}, repeating the process five times. We measure the effectiveness of predicting high-risk option combinations by counting the number of combinations that trigger vulnerabilities during the fuzz tests.}

\newtext{As shown in Table~\ref{tab:ablation}, within these 25 programs, ProphetFuzz identifies 221 high-risk option combinations that triggers vulnerabilities, which is 2.63 times the number identified by ProphetFuzz$^{NSC}$ (84 combinations) and 1.44 times that of ProphetFuzz$^{ZS}$ (153 combinations). This demonstrates that equipping ProphetFuzz with both a self-check mechanism and few-shot methods significantly enhances its performance in predicting high-risk option combinations. ProphetFuzz$^{NSC}$ exhibits the poorest performance, likely due to unchecked constraints disrupting the LLM's reasoning process. For example, in the six programs—\textit{cmark}, \textit{djpeg}, \textit{gif2png}, \textit{jasper}, \textit{pdftohtml}, and \textit{pngfix}—all options are considered conflicting by the LLM, which prevents it from predicting any high-risk option combinations that met the constraints. This underscores the critical importance of the self-check method. In contrast, ProphetFuzz$^{ZS}$, which includes the self-check mechanism, performs better than ProphetFuzz$^{NSC}$ but is less effective than the fully equipped ProphetFuzz, highlighting the efficacy of the few-shot method.}

\subsection{\newtext{Performance of Out-of-the-Box LLMs in Predicting High-Risk Combinations (RQ4)}}
\label{subsec:evaluation_out-of-the-box}

\newtext{To discuss whether an Out-of-the-Box LLM can effectively predict high-risk option combinations, we employ GPT-4 turbo—the same model used in ProphetFuzz—to predict high-risk combinations for the 25 programs selected in Section~\ref{subsec:evaluation_self-check_few-shot}. Specifically, we provide the same program documentation used in ProphetFuzz but employ a straightforward prompt: \textit{``Please predict the high-risk option combinations.''} Note that without our carefully designed prompts, the Out-of-the-Box LLM lacks the capability for command assembly and file generation. To assess its predicted combinations, we process them using our command assembly module and conduct the experiment concurrently with those described in Section~\ref{subsec:evaluation_self-check_few-shot} under the same operational conditions, repeating the process five times.}

\begin{table}[htbp]
    \centering
    \footnotesize
    \caption{\newtext{Results of a 72-hour ablation study on 25 randomly selected programs, evaluating the number of vulnerable combinations identified by each configuration: ProphetFuzz, ProphetFuzz$^{NSC}$ (without the self-check mechanism), ProphetFuzz$^{ZS}$ (without the few-shot method), and OBLLM (Out-of-the-Box LLM with the command assembly module). }}
    \newtext{    
    \begin{threeparttable}
        \begin{tabular}{ccccc}
        \hline
        \textbf{Program} & \textbf{ProphetFuzz} & \textbf{Prophet$^{NSC}$} & \textbf{Prophet$^{ZS}$} & \textbf{OBLLM} \\ \hline
            cflow & \textbf{34} & 18 & 21 & 25 \\ 
            cmark & 0 & -\tnote{*} & 0 & 0 \\ 
            djpeg & 0 & -\tnote{*} & 0 & 0 \\ 
            dwarfdump & \textbf{7} & 1 & 1 & 4 \\ 
            eu-elfclassify & 0 & 0 & 0 & 0 \\ 
            exiv2 & \textbf{5} & 0 & 4 & 3 \\ 
            ffmpeg & \textbf{17} & 3 & 9 & 7 \\ 
            gif2png & 26 & -\tnote{*} & \textbf{27} & 16 \\ 
            gm & 0 & \textbf{1} & 0 & 0 \\ 
            img2sixel & \textbf{24} & 2 & 13 & 2 \\ 
            jasper & 0 & -\tnote{*} & 1 & 0 \\ 
            jpegoptim & 10 & 6 & 8 & \textbf{18}\\ 
            jpegtran & 1 & 1 & \textbf{2} & 0 \\ 
            lrzip & 0 & 0 & 0 & 0 \\ 
            mutool & \textbf{6} & 0 & 0 & 4 \\ 
            nasm & \textbf{2} & 1 & 0 & 1 \\ 
            objdump & \textbf{13} & 8 & 8 & \textbf{13} \\ 
            openssl-ec & 0 & 0 & 0 & 0 \\ 
            openssl-rsa & 0 & 0 & 0 & 0 \\ 
            pdftohtml & 28 & -\tnote{*} & 27 & \textbf{30} \\ 
            pdftopng & \textbf{27} & 17 & 15 & 6 \\ 
            pngfix & 0 & -\tnote{*} & 0 & 0 \\ 
            size & 0 & 0 & 0 & 0 \\ 
            tiffcrop & 19 & \textbf{26} & 13 & 16 \\ 
            xmllint & 2 & 0 & \textbf{4} & 0 \\ \hline
            \textbf{Count} & \textbf{221} & 84 & 153 & 145 \\\hline
        \end{tabular}
    \begin{tablenotes}
        \footnotesize
        \item[*] Fail to predict.
      \end{tablenotes}
    \end{threeparttable}
    \label{tab:ablation}
    }
\end{table}

\newtext{As shown in Table~\ref{tab:ablation}, the Out-of-the-Box LLM identifies 145 combinations that trigger vulnerabilities, showing it also has the ability to predict high-risk combinations. ProphetFuzz outperforms the Out-of-the-Box LLM, triggering vulnerabilities in 1.52 times more combinations, demonstrating greater predictive effectiveness. Notably, ProphetFuzz$^{NSC}$ triggers only about half as many combinations, suggesting that removing the self-check mechanism might detract from the LLM's overall prediction effectiveness. This is likely because unchecked constraints can interfere with reasoning, as discussed in Section~\ref{subsec:evaluation_self-check_few-shot}. Compared to ProphetFuzz$^{ZS}$, which is equipped with self-checked constraints and our CoT prompt but lacks AutoCoT-generated few-shot examples, the Out-of-the-Box LLM performs slightly worse (145 vs 153). This indicates that verified constraints and the CoT prompt enhance the LLM's ability to predict high-risk combinations. Without these enhancements, the Out-of-the-Box LLM might incorrectly identify invalid combinations that violate constraints as high-risk. For instance, in \textit{jpegoptim}, the Out-of-the-Box LLM identifies the combination \textit{``-level1 -level2 -level3''} as high-risk, reasoning that \textit{``they specify different levels of PostScript generation, which may be incompatible or produce conflicting outputs.''} However, despite these advantages, ProphetFuzz$^{ZS}$ does not outperform the Out-of-the-Box LLM as significantly as ProphetFuzz does, suggesting that without AutoCoT-generated few-shot examples, the LLM learns less about predicting high-risk option combinations from the CoT prompt, leading to performance comparable to the Out-of-the-Box LLM.}

\subsection{Contribution of Command Assembly (RQ5)}
\label{subsec:evaluation_command_assembly}

To assess the contribution of the command assembly module to ProphetFuzz's performance, we conduct ablation experiments on the 25 programs randomly selected \newtext{in Section~\ref{subsec:evaluation_self-check_few-shot}.} Specifically, we compare two setups: one in which ProphetFuzz operates without using generated option values (ProphetFuzz$^{NV}$) and another without using generated seed files (ProphetFuzz$^{NS}$). In the ProphetFuzz$^{NV}$ 's setup, we utilize default option values from the dataset along with the seeds generated by ProphetFuzz to assemble the predicted high-risk option combinations. Conversely, ProphetFuzz$^{NS}$ combines the seed input files provided by the dataset with option values generated by ProphetFuzz to assemble commands. \newtext{Since command assembly occurs after prediction, we use end-to-end metrics for evaluation, that is, the number of unique vulnerabilities and edge coverage.} Each setup is run for 72 hours in the same environment and repeated five times. The results are shown in Table~\ref{tab:ablation_command_assembly}.

\begin{table}[htbp]
    \centering
    \footnotesize
    \caption{Results of a 72-hour ablation study on 25 randomly selected programs, \newtext{detailing the number of unique vulnerabilities (\#vuls) and edge coverage (\#cov.) for each configuration:} ProphetFuzz, ProphetFuzz$^{NV}$ (without generated option values) and ProphetFuzz$^{NS}$ (without generated seeds).}
    \begin{tabular}{ccccccc}
    \hline
    \multirow{2}{*}{\textbf{Program}} & \multicolumn{2}{c}{\textbf{ProphetFuzz}} & \multicolumn{2}{c}{\textbf{ProphetFuzz$^{NV}$}} & \multicolumn{2}{c}{\textbf{ProphetFuzz$^{NS}$}} \\ 
        & \#vuls & \#cov & \#vuls & \#cov & \#vuls & \#cov \\ \hline
        cflow & \textbf{17} & \textbf{1453} & 10 & 1367 & 12 & 1445\\
        cmark & 0 & 7000 & 0 & \textbf{7739} & 0 & 7001\\
        djpeg & 0 & \textbf{438} & 0 & 400 & 0 & 416\\
        dwarfdump & 2 & 7767 & \textbf{4} & 6958 & 1 & \textbf{7808}\\
        eu-elfclassify & 0 & \textbf{213} & 0 & \textbf{213} & 0 & 203\\
        exiv2 & \textbf{2} & \textbf{8496} & 1 & 8128 & 1 & 7671\\
        ffmpeg & \textbf{4} & 71198 & 1 & 65551 & 2 & \textbf{71667}\\
        gif2png & \textbf{10} & \textbf{441} & \textbf{10} & 427 & \textbf{10} & 440\\
        gm & 0 & \textbf{10172} & 0 & 9415 & 0 & 8727\\
        img2sixel & 5 & 2207 & 3 & 2136 & \textbf{7} & \textbf{2440}\\
        jasper & 0 & \textbf{3416} & \textbf{1} & 3182 & 0 & 2161\\
        jpegoptim & \textbf{4} & \textbf{244} & 2 & 202 & \textbf{4} & 242\\
        jpegtran & 0 & \textbf{5773} & 0 & 5662 & 0 & 5672\\
        lrzip & 0 & 4069 & 0 & 3941 & 0 & \textbf{4706}\\
        mutool & \textbf{1} & 13118 & 0 & 9601 & \textbf{1} & \textbf{16556}\\
        nasm & \textbf{17} & 7399 & 2 & 6205 & 11 & \textbf{8331}\\
        objdump & \textbf{9} & 11392 & 7 & \textbf{11902} & 6 & 11050\\
        openssl-ec & 0 & \textbf{9474} & 0 & 9150 & 0 & 8420\\
        openssl-rsa & 0 & \textbf{8034} & 0 & 7949 & 0 & 7529\\
        pdftohtml & \textbf{3} & 5635 & \textbf{3} & 5611 & 0 & \textbf{6224}\\
        pdftopng & \textbf{12} & 4531 & 11 & 3653 & 5 & \textbf{4895}\\
        pngfix & 0 & \textbf{1058} & 0 & 988 & 0 & 922\\
        size & 0 & \textbf{3708} & 0 & 3602 & 0 & 2842\\
        tiffcrop & 65 & 6283 & 48 & 6038 & \textbf{67} & \textbf{6555}\\
        xmllint & 2 & \textbf{12609} & \textbf{8} & 11764 & 1 & 11681\\\hline
        \textbf{Count} & 136 & & 101 & & 116 \\\hline
    \end{tabular}
    \label{tab:ablation_command_assembly}
\end{table}

Compared to ProphetFuzz$^{NV}$, ProphetFuzz detects 34.65\% more vulnerabilities and increases edge coverage by 8.21\%, indicating that the option values generated by the command assembly module effectively enhance ProphetFuzz's performance. In contrast, compared to ProphetFuzz$^{NS}$, ProphetFuzz discovers 17.24\% more vulnerabilities but only achieves a slight increase in coverage (0.26\%). This shows that the seed files from the command assembly module markedly improve vulnerability detection but only slightly increase path coverage across the 25 programs. 

In \textit{xmllint}, ProphetFuzz finds notably fewer vulnerabilities than ProphetFuzz$^{NV}$, and the problem arises because the default option values in the dataset incidentally trigger vulnerabilities. The critical option associated with these vulnerabilities is \textit{``--maxmem''}, which is used to specify the maximum number of bytes allocable. We discover that vulnerabilities are triggered only when the option value falls within the range of $[97406, 103252]$, and the default option value incidentally falls within this range ($102400$). However, the values generated by ProphetFuzz are to induce memory-related abnormal behaviors, which are either too small ($<10240$) or too large ($>1000000$), failing to trigger these vulnerabilities.

\subsection{Zero-day Vulnerablities Discovered (RQ6)}
\label{subsec:evaluation_zeroday}

To verify ProphetFuzz's ability to uncover zero-day vulnerabilities, we conduct persistent fuzzing on the latest versions of 52 programs in our evaluation dataset. The discovered vulnerabilities represent unpatched flaws, potentially qualifying as zero-day or half-day vulnerabilities. As shown in Table~\ref{tab:zeroday}, ProphetFuzz identifies 140 unique vulnerabilities across 15 programs, covering 11 types, including heap overflow, stack overflow, use-after-free, and double-free. We report these findings to the developers; \newtext{93} are confirmed, \newtext{26} fixed, and \newtext{21} assigned CVE numbers. Notably, after reporting a use-after-free vulnerability to the \textit{dwarfdump} developers, they express intent to include our ``nice'' test case in their regression tests, underscoring the effectiveness of our command assembly module.

Next, we analyze specific cases to demonstrate how ProphetFuzz successfully identifies these vulnerabilities. 

\noindent\textbf{Bad free in \textit{editcap}}. In \textit{editcap}, a bad-free vulnerability is triggered by combining \textit{``--inject-secrets <secrets type>,<file>''} with \textit{``-c''}. ProphetFuzz extracts constraints from the documentation (e.g., \textit{-c} conflicts with \textit{-i}) to guide the LLM in avoiding invalid combinations. During the prediction process, ProphetFuzz notes, \textit{``Combining the error introduction (-E) with other packet modification operations may lead to unexpected memory states,''} and flags the combination of \textit{--inject-secrets}, \textit{-E}, and \textit{-c} as high-risk. It then assignes values to these options and generates the necessary configuration and input files. Notably, the configuration file for \textit{--inject-secrets} must adhere to the strict format specifications of the NSS key log. Despite this complexity, ProphetFuzz successfully generates the file by leveraging its built-in knowledge. This enables ProphetFuzz to identify the vulnerability during fuzz testing.

\noindent\textbf{Buffer overflow in \textit{jpegtran}}. In \textit{jpegtran}, a buffer overflow is triggered by the \textit{``-drop +X+Y filename''} option, which inserts an image at specified coordinates. After extracting constraints and avoiding invalid option combinations, ProphetFuzz identifies \textit{-drop} with \textit{-maxmemory} as high-risk, noting: \textit{``The -drop option could be sensitive to buffer overflows if the dropped image's dimensions or encoding parameters cause unexpected memory usage.''} Notably, \textit{X} and \textit{Y} represent the coordinates in the input image where another image is to be inserted. If the new image's dimensions exceed those of the input image, the program reports an \textit{``Invalid crop request''} error and exits. Leveraging the LLM's understanding of documentation, ProphetFuzz assigns values of 10 and 20 to \textit{X} and \textit{Y}, and generates a 50x50 configuration file along with a 100x100 seed file to ensure the program could operate correctly. These configurations ultimately help discover this buffer overflow vulnerability.

\begin{table}[htbp]
    \centering
    \footnotesize
    \caption{Vulnerabilities identified by ProphetFuzz. All these vulnerabilities were found in the latest versions of the programs, indicating that they are either zero-day or half-day vulnerabilities.}
    \begin{threeparttable}
    \begin{tabular}{ccccc}
    \hline
    \textbf{Program} & \textbf{Vulnerability Type} & \textbf{Count} & \textbf{\#Fixed} & \textbf{\#CVEs} \\ \hline
        avconv & assertion failure & 1 & 0 & 0\\ 
        avconv & floating point exception & 1 & 0 & 0\\ 
        avconv & global-buffer-overflow & 1 & 0 & 0\\ 
        avconv & heap-buffer-overflow & 21 & 0 & 0\\ 
        avconv & segmentation violation & 6 & 0 & 0\\ 
        avconv & use-after-free & 1 & 0 & 0\\ 
        bison & segmentation violation & 10 & 0\tnote{+} & 0\\ 
        c++filt & stack-buffer-overflow & 12 & 0\tnote{+} & 0\\ 
        cflow & global-buffer-overflow & 1 & 0 & 0\\ 
        cflow & use-after-free & 2 & 0 & 0\\ 
        dwarfdump & use-after-free & 1 & 1 & 1\\ 
        editcap & bad free & 2 & 2 & \newtext{1}\\ 
        editcap & heap-buffer-overflow & 1 & 1 & \newtext{1}\\ 
        editcap & segmentation violation & 1 & 1 & 0\tnote{*}\\ 
        ffmpeg & floating point exception & 2 & \newtext{1\tnote{+}} & \newtext{1}\\ 
        ffmpeg & segmentation violation & 3 & 2 & \newtext{2}\\ 
        img2sixel & floating point exception & 2 & 0 & 0\\ 
        img2sixel & heap-buffer-overflow & 1 & 0 & 0\\ 
        img2sixel & segmentation violation & 1 & 0 & 0\\ 
        img2sixel & use-after-free & 1 & 0 & 0\\ 
        jasper & assertion failure & 1 & 1 & 1\\ 
        jpegtran & segmentation violation & 1 & \newtext{1} & 0\tnote{*}\\ 
        mupdf & negative-size-param & 1 & 1 & 1\\ 
        mupdf & segmentation violation & 1 & 1 & 1\\ 
        nasm & segmentation violation & 4 & 0 & 0\\ 
        nasm & use-after-free & 3 & 0 & 0\\ 
        objdump & heap-buffer-overflow & 2 & 1 & 1\\ 
        pspp & assertion failure & 29 & \newtext{8\tnote{+}} & \newtext{6}\\ 
        pspp & bus on unknown address & 1 & 0\tnote{+} & 0\\ 
        pspp & double-free & 1 & 0\tnote{+} & 0\\ 
        pspp & heap-buffer-overflow & 3 & 0\tnote{+} & 0\\ 
        pspp & segmentation violation & 13 & \newtext{4\tnote{+}} & \newtext{4}\\ 
        pspp & stack-buffer-overflow & 5 & 0\tnote{+} & 0\\ 
        pspp & use-after-free & 3 & 0\tnote{+} & 0\\ 
        xpdf & stack-buffer-overflow & 1 & 1 & 1 \\ \hline
        \textbf{Count} & & 140 & \newtext{26} & \newtext{21}\\ \hline
    \end{tabular}
    \begin{tablenotes}
        \footnotesize
        \item[+] Have been confirmed by the developers.
        \item[*] Applying for CVE.
      \end{tablenotes}
    \end{threeparttable}
    \label{tab:zeroday}
\end{table}

\subsection{Predictive Knowledge (RQ7)}
\label{subsec:evaluation_knowledge}

To understand how ProphetFuzz, without any expert intervention, manages to predict 1748 high-risk option combinations solely from program documentation, we record the complete CoT outputs from the LLM during each prediction. Initially, we extract content directly related to the predictions, focusing on the fourth step of the prompt. We then instruct the LLM to \textit{``summarize the knowledge of the categories of program option combinations that may have potential vulnerabilities.''} Subsequently, we compile summaries for each program, have the LLM synthesize these summaries and rank them from most to least common. To make the results more understandable, we set the context in the prompt: \textit{``For software testers who do not understand security, they hope to have easy-to-understand knowledge to conduct efficient security testing.''} Finally, we capture 15 key pieces of knowledge crucial for predicting high-risk option combinations. The top three are detailed below (see footnote~\ref{fn:repo} for a complete list).

\textbf{1. Resource Management and Limits}. Options that affect resource allocation and limits can lead to vulnerabilities when they conflict with options that increase resource demands, potentially causing resource exhaustion or buffer overflows.

\textbf{2. Complex Data Processing}. Combinations of options that lead to complex data processing tasks can increase the risk of vulnerabilities such as memory corruption, especially when involving external data or detailed output formatting.
    
\textbf{3. Output and Format Manipulation}. Options that modify output verbosity or format can lead to vulnerabilities if they result in excessive data being processed or displayed, potentially revealing sensitive information or causing buffer overflows.

\ifthenelse{\boolean{showrevisions}}
    {\section{\removedtext{Discussion}\newtext{Limitation and Future Work}}}
    {\section{Limitation and Future Work}}
\label{sec:discussion}

\ifthenelse{\boolean{showrevisions}}
    {\subsection{\removedtext{Performance of the LLM with Basic Prompts}}}
    {}

\removedtext{To assess the effectiveness of LLMs without our carefully designed prompts used by ProphetFuzz, we conduct experiments using the GPT-4 Turbo model and explore what performance can be achieved by the LLM when more straightforward prompts are applied. We provide the same program documentation used in ProphetFuzz but employ a straightforward prompt: \textit{``Please predict the high-risk option combinations and assemble executable commands for these high-risk option combinations.''} This approach reveals several issues with the results, which are discussed below:}

\removedtext{Without our prompts, the LLM mistakenly identifies invalid option combinations that violate constraints as high-risk instead of those likely to cause deep memory vulnerabilities. For instance, in \textit{jpegoptim}, the LLM considers the combination of \textit{-level1}, \textit{-level2}, and \textit{-level3} to be high-risk, reasoning that \textit{``they specify different levels of PostScript generation, which may be incompatible or produce conflicting outputs.''} This misjudgment arises from, on the one hand, the LLM lacking specific guidance to avoid invalid combinations. On the other hand, the LLM lacks the knowledge necessary to analyze high-risk combinations, leading to imprecise predictions. Moreover, without additional guidance, while the LLM can successfully assemble the predicted option combinations, it does not generate corresponding seed files autonomously. Typically, seed files are manually collected from the internet, but this can lead to semantic mismatches between the seed files and option values. ProphetFuzz, by integrating constraint extraction, high-risk option combination prediction, and command assembly modules, effectively resolves these issues and enhances performance.}

\ifthenelse{\boolean{showrevisions}}
    {\subsection{\removedtext{Limitation and Future Work}}\label{subsec:discussion_limitation_future_work}}
    {}

While ProphetFuzz demonstrates impressive performance in our evaluation experiments, it still has some limitations. In some programs, the precision of constraint extraction by ProphetFuzz is unsatisfactory primarily because its constraint extraction module heavily relies on the LLM's reasoning performance. When the LLM underperforms, ProphetFuzz's performance is correspondingly affected. With the continuous advancements in LLM technology, we expect ProphetFuzz's performance to improve. \newtext{Additionally, ProphetFuzz solely relies on documentation for predictions, which is both a strength and a limitation. When documentation is incorrect or missing, it becomes difficult for ProphetFuzz to perform constraint extraction, target prediction, or command assembly. In the future, we aim to enable ProphetFuzz to reference additional sources of information, such as code comments and the code itself. Moreover, despite ProphetFuzz achieving full automation in predicting and fuzzing high-risk option combinations, the historical high-risk option combinations used to generate few-shot corpora are manually collected. Note that it is a one-time job, and we have made all these combinations publicly available, ensuring that the process does not need to be repeated.} 
For complex multi-option constraints, the ideal approach is to split them into pairwise option constraints for self-checks, which provides the highest precision but would significantly increase the number of LLM queries. For cost considerations, ProphetFuzz instructs the LLM to evaluate the entire multi-option constraint in a single inference. While this approach is more cost-effective, it may confuse the LLM's analysis process and introduce incorrect constraints.

In a few programs, the high-risk option combinations predicted by ProphetFuzz do not achieve satisfactory results. On the one hand, this is because ProphetFuzz primarily focuses on mutating input files, whereas for some programs (e.g., \textit{vim}), mutating configuration files might yield better outcomes. To address this, we plan to enable the LLM to autonomously choose the most suitable target for mutation in future versions. On the other hand, although we design an automated few-shot corpus collection method that generates the analysis examples for predicting some high-risk option combinations without expert involvement, real expert experience may still provide more effective learning outcomes for the LLM. Therefore, we intend to involve security experts to conduct a deeper analysis of the prediction of high-risk option combinations, aiming to produce a higher-quality corpus. 

Additionally, ProphetFuzz currently focuses only on predicting high-risk option combinations that lead to memory crash vulnerabilities, aligning with the characteristics of fuzzers. By adjusting the prompts, ProphetFuzz can also be adapted to predict other types of vulnerabilities, such as multithreading and undefined behavior errors, in future implementations.

\section{Conclusion}
\label{sec:conclusion}

This paper introduces ProphetFuzz, a fully automated tool based on large language models (LLM) specifically designed to predict high-risk option combinations and conduct fuzz testing. Benefiting from our thoughtfully engineered prompts and the LLM's superior text comprehension and reasoning capabilities, ProphetFuzz successfully predicts 1748 high-risk option combinations on a dataset covering 52 programs, leading to the discovery of 364 unique vulnerabilities, which is 32.85\% more effective than the previous work. Additionally, ProphetFuzz identifies 140 zero-day or half-day vulnerabilities in the latest versions of these programs, with \newtext{93} confirmed by developers and \newtext{21} awarded CVE numbers.
\section*{Acknowledgement}

We are grateful to our shepherd and the anonymous reviewers for their valuable guidance and insightful comments.

\bibliographystyle{ACM-Reference-Format}
\balance
\bibliography{references}
\appendix
\section*{Appendix}

\renewcommand{\thesubsection}{\Alph{subsection}}

\subsection{Prompts for Code Assembly}
\label{appendix:prompt_code_assembly}

\begin{figure*}[htbp]
    \centering
    \includegraphics[width=0.8\textwidth]{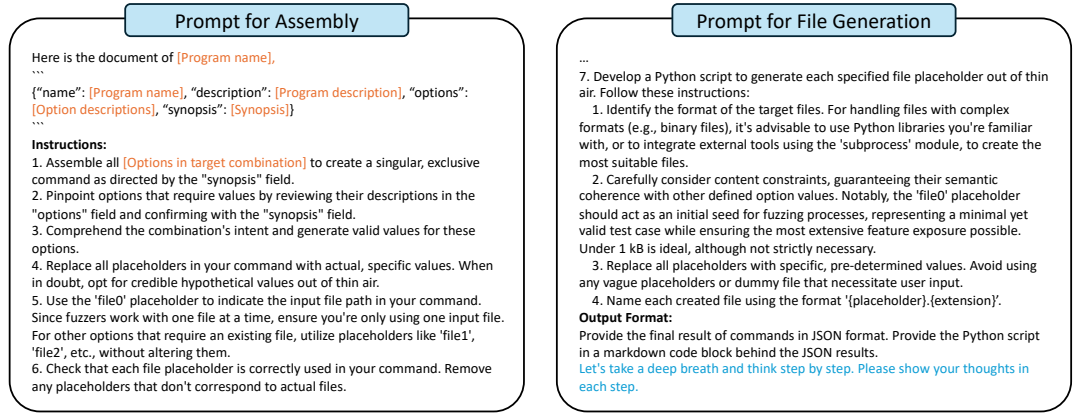}
    \caption{Prompts for assembly and file generation. The content enclosed in brackets denotes the need for specific input.}
    \label{fig:assembly_prompt}
\end{figure*}

Figure~\ref{fig:assembly_prompt} shows the prompt for code assembly.

\subsection{\newtext{Recall for 20 Randomly Selected Programs}}
\label{appendix:recall}

\begin{table}[htbp]
    \centering
    \caption{\newtext{Precision (\textit{Prec.}) and Recall (\textit{Rec.}) for Constraint Extraction by ProphetFuzz, CarpetFuzz, and ProphetFuzz$^{NSC}$ from Documentation of 20 Randomly Selected Programs.}}
    \footnotesize
    \newtext{
    \begin{tabular}{ccccccc}
    \hline
    \multirow{2}{*}{\textbf{Program}} & \multicolumn{2}{c}{\textbf{ProphetFuzz}} & \multicolumn{2}{c}{\textbf{CarpetFuzz}} & \multicolumn{2}{c}{\textbf{ProphetFuzz$^{NSC}$}} \\ 
        & Prec. & Rec. & Prec. & Rec. & Prec. & Rec. \\ \hline
    cjpeg & 100.0\% & 80.0\% & 50.0\% & 40.0\% & 2.1\% & 100.0\%\\ 
    cmark & 100.0\% & 50.0\% & 100.0\% & 50.0\% & 11.1\% & 100.0\%\\ 
    djpeg & 100.0\% & 74.3\% & 95.5\% & 60.0\% & 7.3\% & 77.1\%\\ 
    exiv2 & 100.0\% & 80.0\% & 50.0\% & 20.0\% & 8.5\% & 80.0\%\\ 
    gif2png & 100.0\% & 100.0\% & 0.0\% & 0.0\% & 1.4\% & 100.0\%\\ 
    img2sixel & 87.5\% & 87.5\% & 100.0\% & 62.5\% & 9.2\% & 100.0\%\\ 
    lrzip & 100.0\% & 100.0\% & 100.0\% & 41.2\% & 7.6\% & 100.0\%\\ 
    openssl-asn1parse & 100.0\% & 25.0\% & 100.0\% & 50.0\% & 4.0\% & 100.0\%\\ 
    openssl-ec & 100.0\% & 100.0\% & 100.0\% & 60.0\% & 7.5\% & 100.0\%\\ 
    pdftopng & 100.0\% & 50.0\% & 20.0\% & 50.0\% & 66.7\% & 100.0\%\\ 
    pdftotext & 100.0\% & 94.1\% & 22.2\% & 11.8\% & 6.6\% & 100.0\%\\ 
    podofoencrypt & 100.0\% & 100.0\% & 100.0\% & 33.3\% & 14.3\% & 100.0\%\\ 
    pspp & 100.0\% & 77.8\% & 0.0\% & 0.0\% & 11.2\% & 100.0\%\\ 
    size & 90.0\% & 69.2\% & 100.0\% & 92.3\% & 90.9\% & 76.9\%\\ 
    speexdec & 100.0\% & 100.0\% & 45.5\% & 50.0\% & 19.2\% & 100.0\%\\ 
    tcpprep & 100.0\% & 88.9\% & 100.0\% & 77.8\% & 100.0\% & 88.9\%\\ 
    tiff2ps & 87.5\% & 58.3\% & 62.5\% & 41.7\% & 33.3\% & 58.3\%\\ 
    tiffinfo & 100.0\% & 100.0\% & 0.0\% & 0.0\% & 33.3\% & 100.0\%\\ 
    xmlwf & 100.0\% & 70.0\% & 0.0\% & 0.0\% & 8.1\% & 100.0\%\\ 
    yara & 100.0\% & 100.0\% & 100.0\% & 100.0\% & 2.1\% & 100.0\%\\\hline
    \textbf{Overall} & 97.97\% & 80.56\% & 77.78\% & 46.67\% & 8.56\% & 89.44\% \\\hline
    \end{tabular}
    \label{tab:recall}
    }
\end{table}

\newtext{Table~\ref{tab:recall} shows the precision and recall for Constraint Extraction by ProphetFuzz, CarpetFuzz, and ProphetFuzz$^{NSC}$ from Documentation of 20 Randomly Selected Programs.}


\subsection{Complete List of Predictive Knowledge}
\label{appendix:knowledge}

The complete list of the extracted 15 pieces of knowledge used by ProphetFuzz to predict high-risk option combinations is as follows:

\begin{enumerate}[1.]
    \item \textbf{Resource Management and Limits}. Options that affect resource allocation and limits can lead to vulnerabilities when they conflict with options that increase resource demands, potentially causing resource exhaustion or buffer overflows.

    \item \textbf{Complex Data Processing}. Combinations of options that lead to complex data processing tasks can increase the risk of vulnerabilities such as memory corruption, especially when involving external data or detailed output formatting.
    
    \item \textbf{Output and Format Manipulation}. Options that modify output verbosity or format can lead to vulnerabilities if they result in excessive data being processed or displayed, potentially revealing sensitive information or causing buffer overflows.
    
    \item \textbf{Error Handling Modifications}. Options that suppress or alter error handling can hide underlying issues, allowing the program to operate in an unstable state and increasing the risk of vulnerabilities.
    
    \item \textbf{Conflicting Operations}. Using options that perform opposing actions can lead to undefined behavior or race conditions, potentially causing the software to enter an unstable state.
    
    \item \textbf{Input/Output Handling}. Options that affect how input and output are handled can lead to vulnerabilities if they cause the program to read or write outside of intended memory areas or handle file operations insecurely.
    
    \item \textbf{Concurrency and Parallel Processing}. Options that enable multi-threading or parallel processing can introduce vulnerabilities such as race conditions if combined with options that are not thread-safe.
    
    \item \textbf{Verbose and Debugging Modes}. Increasing the verbosity of the program's output or enabling debugging modes can inadvertently expose vulnerabilities by providing more data to an attacker or changing the timing and performance characteristics of the application.
    
    \item \textbf{Security Thresholds and Protections}. Options that set security thresholds can lead to vulnerabilities when incorrectly configured or combined with complex XML structures, as they may not protect sufficiently or cause legitimate processing to fail.
    
    \item \textbf{External Data and Variable Definition}. Allowing external data input or variable definition can lead to vulnerabilities when combined with options that do not properly sanitize or handle this external input.
    
    \item \textbf{Transformation and Canonicalization}. Transforming input into detailed or canonical forms can increase the complexity of processing, leading to vulnerabilities due to the increased complexity of the output.
    
    \item \textbf{Specialized Processing Modules}. Passing data to specialized processing modules can lead to vulnerabilities if the modules do not properly handle the data, especially when combined with options that modify data handling.
    
    \item \textbf{Scan Optimization}. Optimizing scanning for performance can lead to vulnerabilities when it conflicts with thorough scanning required for security, potentially missing critical checks.
    
    \item \textbf{Memory-Intensive Operations}. Combining options that are inherently memory-intensive can lead to vulnerabilities, especially if they are not properly optimized for memory usage, potentially resulting in memory leaks or corruption.
    
    \item \textbf{Control Flow Alteration}. Changing the program's control flow with certain options can lead to vulnerabilities when combined with extensive processing options, potentially leading to incorrect processing or logic bypass.
\end{enumerate}

\end{document}